\documentclass{amsart}
\usepackage{geometry}                
\usepackage[parfill]{parskip}   
\usepackage{graphicx}
\graphicspath{ {./figures/} }
\usepackage{amssymb}
\usepackage{epstopdf}
\usepackage{caption}
\usepackage{subcaption}
\usepackage{lscape}
\usepackage{physics}
\usepackage{hyperref}
\usepackage{amsaddr}
\usepackage{array}   
\newcolumntype{L}{>{$}c<{$}} 
\usepackage{endnotes}
\begin{document}

\title{Dense Suspension Flow in a Penny-Shaped Crack \\ Part I : Theory}
\author{George R. Wyatt\textsuperscript{1} \& Herbert E. Huppert\textsuperscript{2}}
\address[1]{\textsuperscript{1}Emmanuel College, St. Andrew's Street, Cambridge, CB2 3AP.  grw44@cam.ac.uk \\ \textsuperscript{2}King's College, King's Parade, Cambridge, CB2 1ST.  heh1@cam.ac.uk}
\keywords{Hydraulic fracture, suspension flow, rheology, proppant transport, elastic, tip screen-out, penny-shaped, cavity flow}

\begin{abstract}
We study the dynamics of proppants carried by fluid driven into an evolving penny-shaped fracture. The behaviour of the slurry flow is investigated in two phases: pressurised injection and elastic closure. During injection the slurry is modelled using a frictional rheology that takes into account the shear-induced migration and jamming of the proppants. Making pragmatic assumptions of negligible toughness and cross-fracture fluid slip, we find self-similar solutions supporting a range of proppant concentration profiles. In particular, we define an effective viscosity, which equates the fracture evolution of a slurry flow with a given proppant volume fraction, to a Newtonian flow with a particular viscosity. Using this framework, we are able to make predictions about the geometry of the growing fracture and the significance of tip screen-out. In the closure phase, proppants are modelled as incompressible and radially immobile within the narrowing fracture. The effects of proppant concentration on the geometry of the residual propped fracture are explored in full. The results have important applications to industrial fracking and geological dike formation by hot, intruding magma.
\end{abstract}

\today
\maketitle

\section{Introduction}
Receiving a patent for his `exploding torpedo' in 1865, US Civil War veteran Col. Edward Roberts established the practice of fracturing bedrock to stimulate oil wells \cite{history1}. A technique, known as hydraulic fracturing, which uses pressurised fluid rather than explosives to develop fracture networks, only came into practice much later, in 1947 \cite{history2}, and is the topic of this paper. In particular, we will concentrate on the convective transport of proppants within an evolving cavity. These are small particles added to the fracturing fluid in order to prop open the developed fracture, which closes under far-field stress once the fluid pressure is released. Aside from its use in hydrocarbon recovery, hydraulic fracturing, or fracking, has uses including the measurement of in-situ stresses in rocks \cite{stressmeasure}, generation of electricity in enhanced geothermal systems \cite{geothermal} and improvement of injection rates in CO$_2$ sequestration \cite{CO2}. Hydraulic fracturing processes are also ubiquitous in geology: dikes and sills arise from cracks whose growth is driven by magma, with magmatic crystals taking the place of synthetic proppants. Phenomena such as crystallisation and gas exsolution in the cooling magma mean models of dike propagation vary widely, as is summarised in \cite{dikesummary}. Notably, Petford \& Koenders \cite{magmatic} utilise granular flow theory to model the ascent of a granitic melt containing solids. 

This paper combines two significant, but often disconnected, fields of fracking study, cavity flow and suspension flow:

\begin{itemize}
\item
The study of (elastohydrodynamic) cavity flow focusses on the interplay between hydrodynamic properties of the fracturing fluid and material properties of the medium being fractured. In the zero-proppant case, the problem of a fluid-driven, penny-shaped crack requires the joint solution of a nonlinear Reynold's equation, which governs flow within the crack, and a singular integral boundary condition, which takes into account the elastic properties of the surrounding medium. The general strategy used in this paper takes inspiration from the work of Spence \& Sharp \cite{1985}, who in 1985, restricting to the two-dimensional case, were the first to solve these integro-differential equations. In particular, we will focus on cavities that keep the same shape in some evolving coordinate system, using series expansions to represent both the width and pressure profiles within the fracture. More recently, in 2002, Savitski \& Detournay \cite{detsav} solved similar three-dimensional versions of these equations, allowing them to find fracture evolutions with simple time dependence in both the viscous and toughness dominated regimes. In the former, the principal energy dissipation is by viscous flow, and in the latter, energy dissipation is mostly by creating new fracture surfaces. Notably, the same paper \cite{detsav} verifies that industrial fracking occurs in the viscous regime; this assumption makes the problem considered in this paper tractable to a semi-analytical approach.

\item
The mathematical study of suspension flow dates back to 1906, when Einstein used properties of suspensions to estimate the size of a water molecule \cite{einstein}. In particular, he showed that very dilute particle-laden flows are Newtonian, with a viscosity which increases with the concentration of particles. However, during hydraulic fracturing it is necessary to model a full range of proppant volume fractions, which we denote by $\phi$. It is typical to have both dilute flow near the crack walls, as well as plug flow at the centre of the cavity, where the slurry behaves as a porous granular medium. More recent experiments by Boyer et al. in 2011 \cite{boyer} investigate dense suspension rheology. They show that particles in suspension, subject to a constant normal particle pressure that is applied by a porous plate, expand when a shear is applied to the mixture. As a result, it is possible to write $\phi=\phi(I)$, where the dimensionless parameter, $I$, is the ratio between the \textit{fluid} shear stress, which is proportional to the shear rate, and the particle normal stress. Likewise, fixing the solid volume fraction, they showed that the normal particle pressure is proportional to the mixture shear stress. It is also shown that the constant of proportionality, $\mu$, can be expressed as a decreasing function of $\phi$. In the same paper \cite{boyer}, forms of the rheological functions $I$ and $\mu$ are suggested, showing good agreement with experimental data. Since then, several papers have suggested slightly different rheological models and are reviewed by Donstov et al. in \cite{comparative}. These all feature a jamming limit, $\phi_m$, which is the volume fraction at which the flowing slurry transitions into a granular solid. We will utilise the frictional rheology given by Lecampion \& Garagash \cite{lecgara}, which is unique in allowing packings with $\phi>\phi_m$. These denser packings form due to ‘in-cage’ particle rearrangements caused by velocity and pressure fluctuations in the surrounding flow.
\end{itemize}
 
The endeavours of this paper may be condensed into three main objectives. The first is to establish a mathematical framework that captures the behaviour of the proppant suspension as it interacts with the growing cavity. Here we will utilise a lubrication model, along with the assumption that the proppant flow is fully developed; equivalently, that the transverse fluid slip is negligible. Crucially, we will try to justify these assumptions using typical parameters from industrial fracking. We will also make a zero-toughness assumption, which is validated in \cite{detsav}.  Once we have developed this framework, an important step will be to compare its features to those derived in the zero-proppant, viscosity dominated case by Savitski \& Detournay \cite{detsav}, particularly because we utilise a frictional rheology fitted to the dense regime. The second objective is to find and examine accurate numerical solutions modelling the developing cavity, given a range of proppant concentrations. We will explore the empirical effects of changing proppant concentration on the geometry of the developing fracture, as well as the distribution of proppants. Where possible, we will evaluate the consistency of our model and forecast potential shortfalls such as proppant screen-out near the crack tip. The third, and final, objective is to leverage our results to make predictions about the geometry of the fracture after the fluid pressure is released. By assuming the remaining proppants are immobile and incompressible, we aim to establish simple formulae predicting the width and radius of the developed fracture. Since these relate directly to the conductivity of the formation, this third objective is potentially the most significant.

Aside from the availability of semi-analytical solutions, the problem of proppant flow in a penny-shaped crack is particularly appealing because of the potential of practical verification. Recent experiments by O’Keeffe, Huppert \& Linden \cite{herbertpractical} have explored fluid-driven, penny-shaped fractures in transparent, brittle hydrogels, making use of small particle concentrations to measure in-crack velocities. This paper is the first of two; the second of which will be a practical treatise on slurry driven-fractures in hydrogels, aiming to verify the predictions made here by repeating the experiments of \cite{herbertpractical} including proppant concentrations.

\newpage
\section{Injection: Problem Formulation}

\begin{figure}[h]
\includegraphics[width=0.8\textwidth]{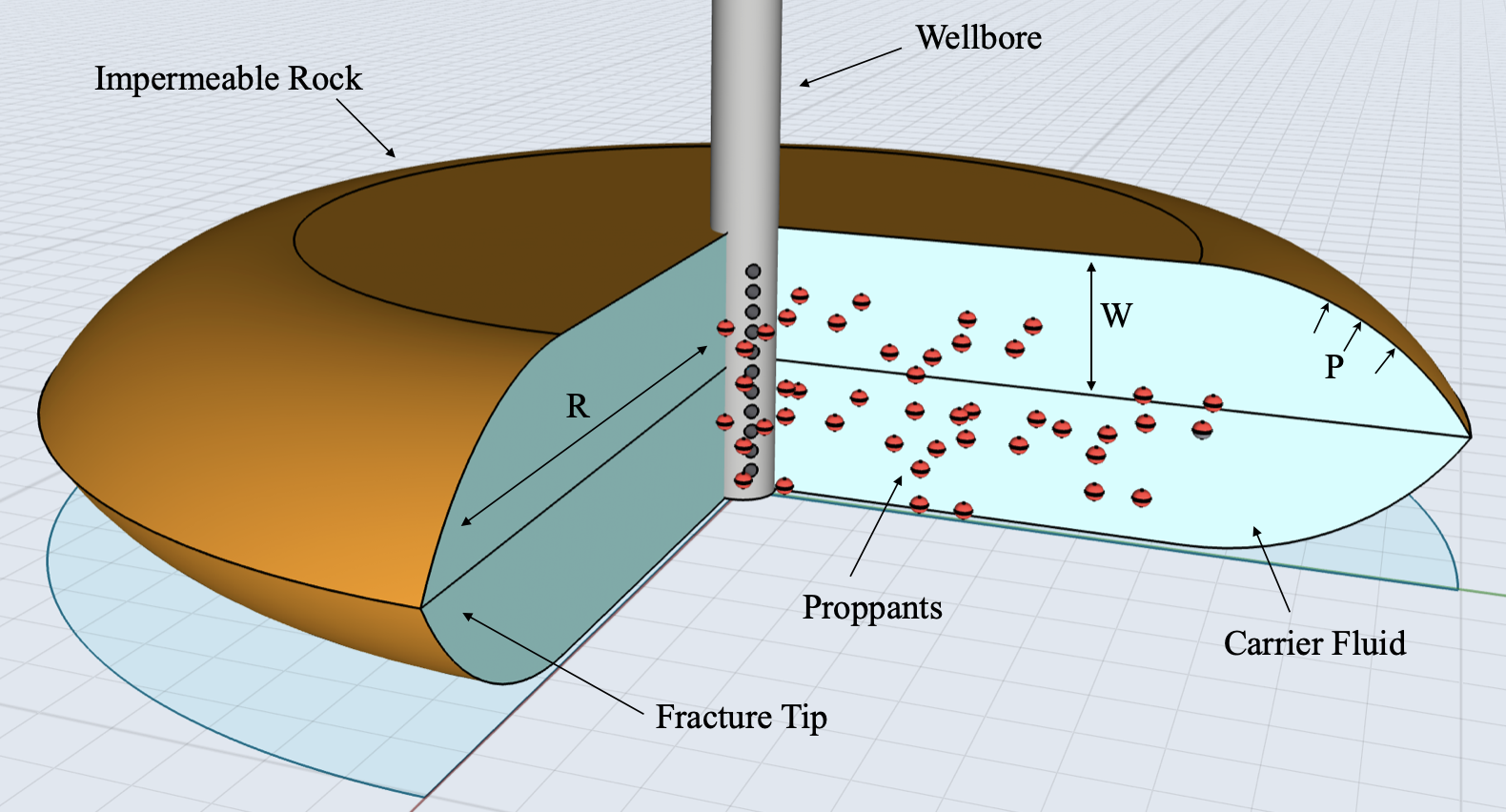}
\centering
\caption{Schematic of the penny-shaped crack.}
\label{schematic1}
\end{figure}
\subsection{Fracture Mechanics}
We model the propagation of a penny-shaped crack similar to that shown in Figure \ref{schematic1}, using the framework of Detournay \& Savitski \cite{detsav}. We will make the following assumptions:
\begin{itemize}
\item The crack is axisymmetric and has reflectional symmetry in $z=0$, with half width $w(r,t)$ and total radius $R(t)$, so $w(R,t)=0$.
\item The fluid is injected from a point source, with the wellbore radius negligible compared to the fracture radius.
\item The lag between the fracture tip and the fluid front is negligible compared to the fracture radius.
\item The fracture propagates in continuous mobile equilibrium.
\item The normal stress on the fracture walls due to proppants is negligible compared to the fluid pressure.
\end{itemize}
The third assumption is validated by Garagash \& Detournay \cite{garadet} and introduces a negative pressure singularity at the tip of the crack ($r=R$).
The fourth and fifth assumptions lead to the following integral equations from linear elastic fracture mechanics. These relate the net fluid  pressure, $p(r,t)$, to the opening of the fracture and the toughness of the surrounding rock.
\begin{align}
w(r,t)&=\frac{4R}{\pi E'} \int _{r/R}^{1} \frac{y}{\sqrt{ y^2-(r/R)^2}} \int_0 ^1 \frac{x p(x y R, t)}{\sqrt{1-x^2}} dxdy, \label{snedoneqn}  \\
K_{Ic}&=\frac{2}{\sqrt{\pi R}} \int_0^R \frac{p(r,t)r}{\sqrt{R^2-r^2}}dr,
\end{align}
where $E'$ is the plane strain modulus, given by the Young modulus, $E$, and the Poisson ratio, $\nu$, as $E'=E/(1-\nu^2)$. $K_{Ic}$ is the material toughness. These equations can be attributed to Sneddon \cite{sneddon} and Rice \cite{rice} respectively. We note that $p$ represents the fluid pressure minus the in-situ stress of the surrounding rock, which is assumed to be isotropic. We write $p$ with radial spatial dependence only; this will be validated later, along with the fifth assumption, using a lubrication argument.


\subsection{Frictional Rheology}
We model the injected flow as a Newtonian fluid containing identical spherical particles. Recent approaches in modelling dense slurry flow are characterised by empirical relations originally proposed by Boyer et al. \cite{boyer}. The first of these relates the fluid shear stress to the normal stress required to confine the particles; the second gives the ratio of the mixture shear stress to the particle confining stress,
\begin{align}
I(\phi)&=\eta_f \dot{\gamma}/\sigma_n^s, & \mu(\phi)&=\tau/\sigma_n^s.
\end{align}
Here $\eta_f$ is the carrying fluid's dynamic viscosity, $\phi$ is the volume fraction of the proppants, $\dot{\gamma}$ is the solid shear rate and $\sigma_n^s$ is the normal particle stress, which we will sometimes refer to as the particle pressure. The second ratio is given the symbol $\mu$, not to be confused with dynamic viscosity, because it resembles a friction coefficient. These relations are given a clear experimental grounding in \cite{boyer}, which is discussed in the introduction. Various forms of the dimensionless functions $I(\phi)$ and $\mu(\phi)$ have been compared to experimental results in \cite{comparative} using the equivalent formulation: $\tau=\eta_s(\phi)\eta_f \dot{\gamma}$ and $\sigma_n=\eta_n(\phi)\eta_f\dot{\gamma}$, where $\eta_s=\mu(\phi)/I(\phi)$ and $\eta_n=1/{I(\phi)}$. 

\begin{figure}
\begin{subfigure}{0.49\textwidth}
\includegraphics[width=\textwidth]{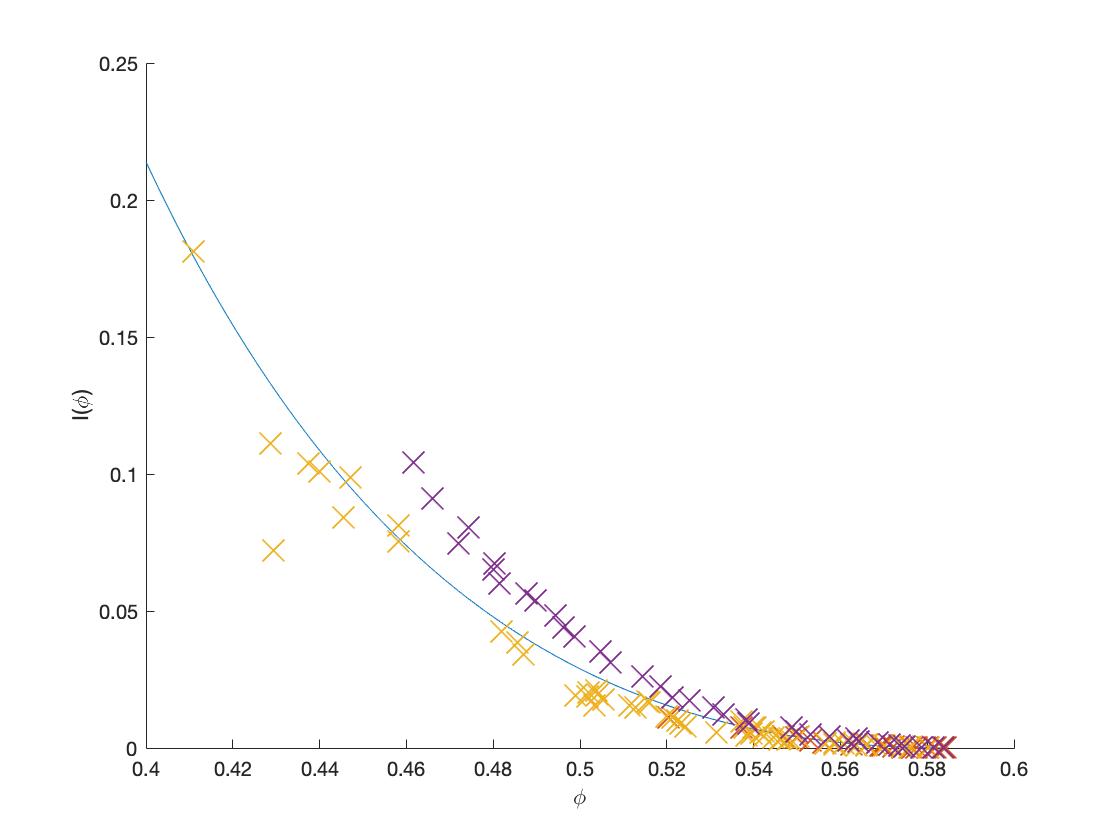}
\caption{$I$}
\end{subfigure}
\begin{subfigure}{0.49\textwidth}
\includegraphics[width=\textwidth]{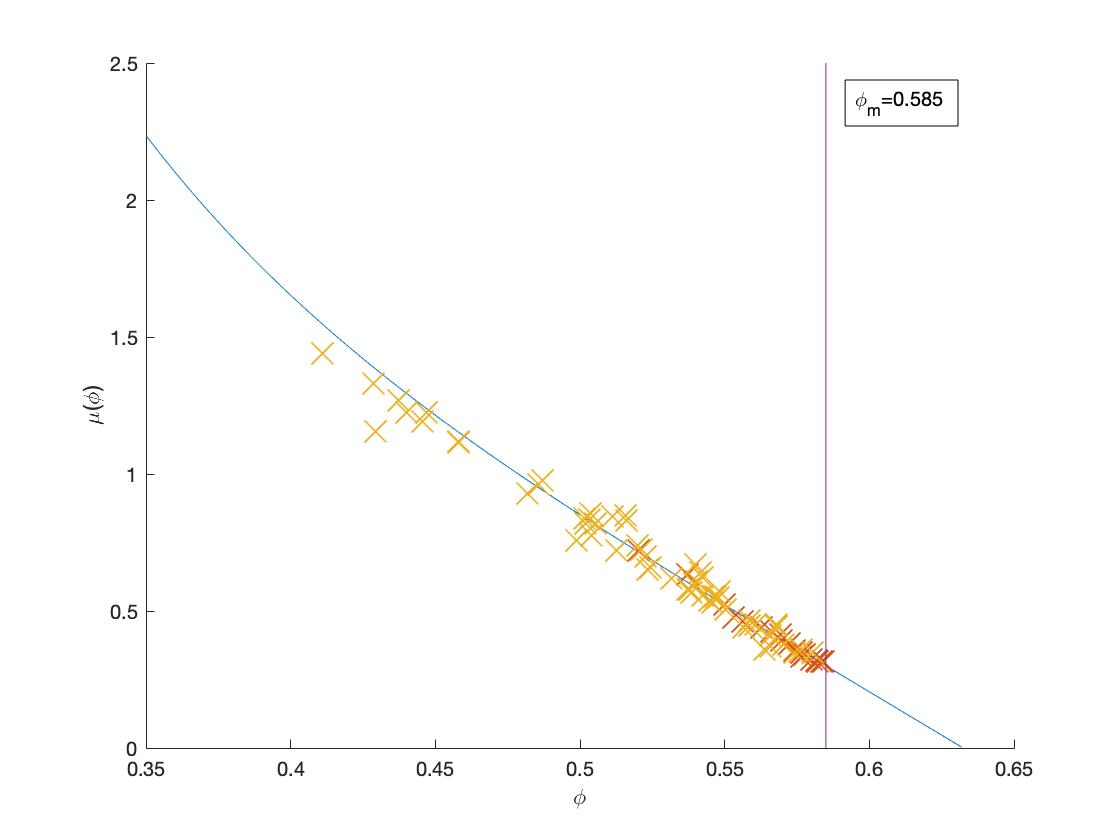}
\caption{$\mu$}
\end{subfigure}
\begin{subfigure}{0.49\textwidth}
\includegraphics[width=\textwidth]{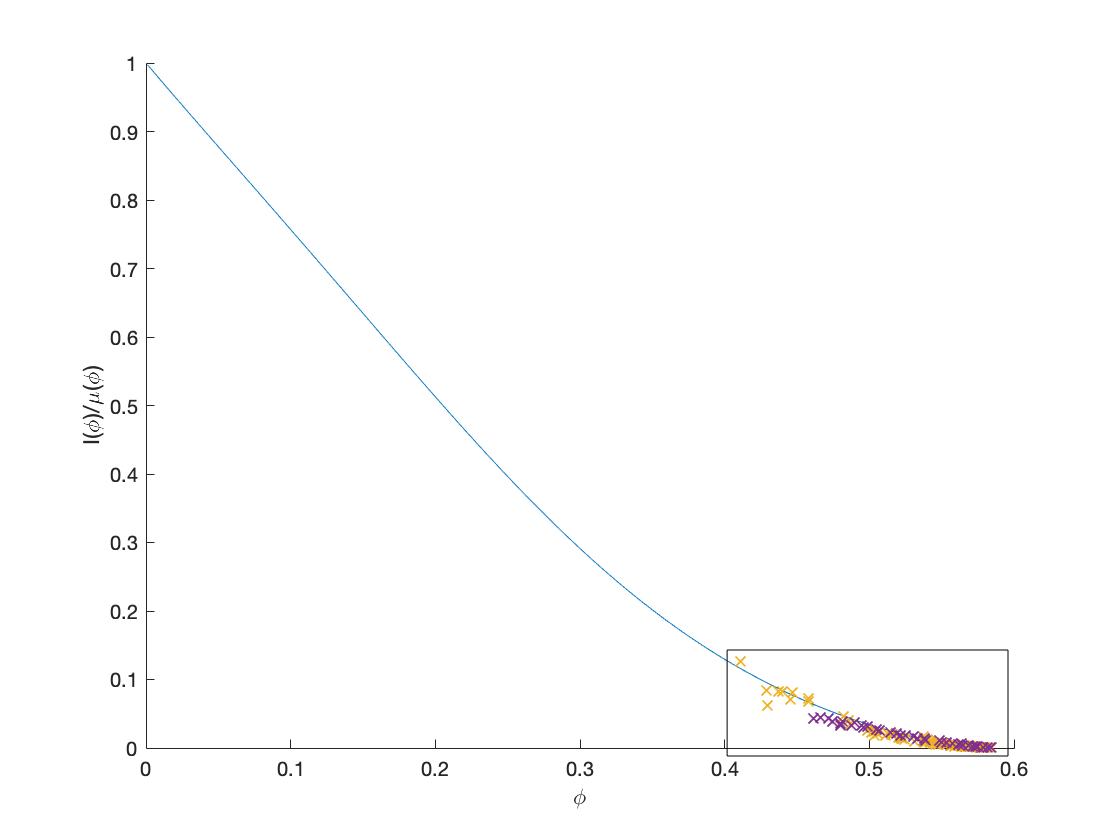}
\caption{$I/\mu$}
\end{subfigure}
\begin{subfigure}{0.49\textwidth}
\includegraphics[width=\textwidth, height=0.75\textwidth]{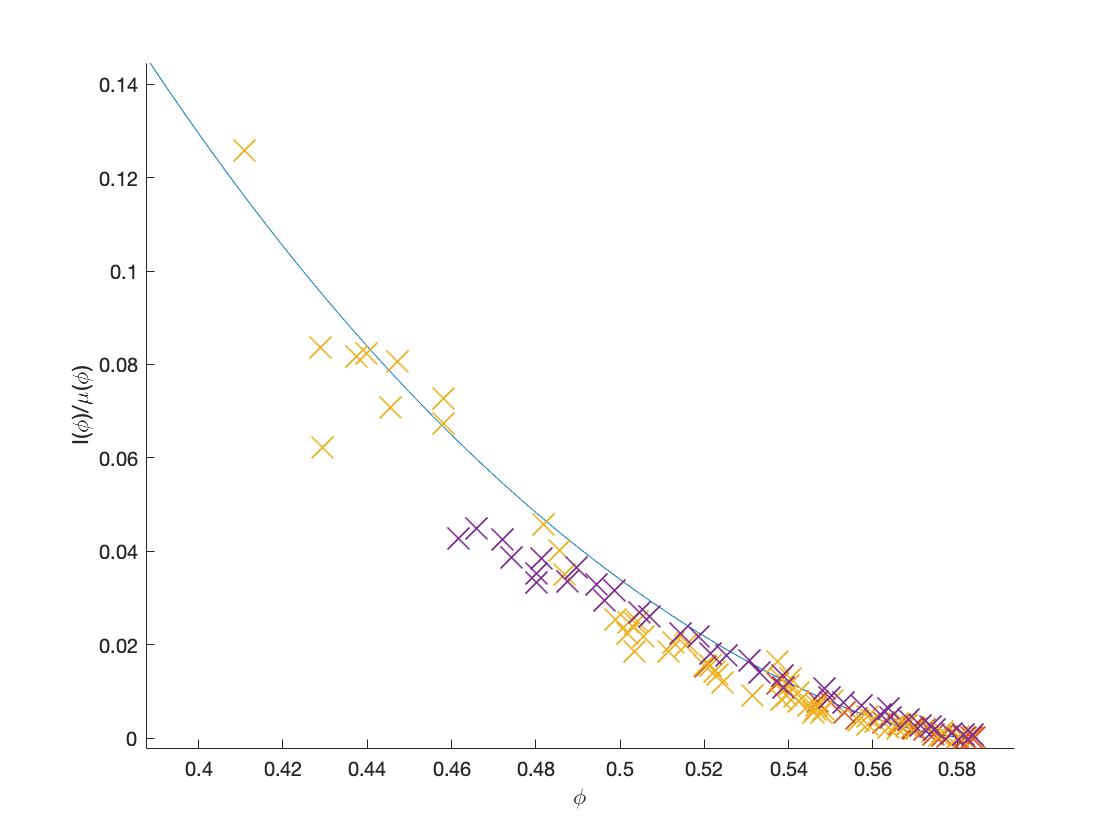}
\caption{$I/\mu$ data}
\label{permeability}
\end{subfigure}
\centering
\caption{Plots of the rheological functions $I$, $\mu$ and $I/\mu$ given by Lecampion \& Garagash \cite{lecgara}. Also plotted is the experimental data of Boyer et al. \cite{boyer} using polystyrene spheres of diameter 580$\mu$m in $2.15$Pa s fluid (red), as well as poly(methyl methacrylate) spheres of diameter 1100$\mu$m suspended in $3.10$Pa s fluid (orange); and of Dagois-Bohy et al. \cite{dagois} using polystyrene spheres of diameter 580$\mu$m suspended in $2.27$Pa s fluid (purple). All experiments are carried out with a fixed particle pressure, applied by a porous plate.} 

\label{rheoplots}
\end{figure}

In our calculations we will utilise the frictional rheology provided by B. Lecampion \& D. I. Garagash \cite{lecgara}, which is unique in allowing packings with volume concentrations greater than $\phi_m$. Here $I(\phi)=0$, meaning the proppants have zero shear rate and effectively resemble a permeable solid. Explicitly, we use the expressions
\begin{align}
\mu=\mu_1+\frac{\phi_m}{\delta}\left(1-\frac{\phi}{\phi_m}\right)&+\left(I(\phi)+\left[\frac{5}{2}\phi_m+2\right]I(\phi)^{0.5}\right)\left(1-\frac{\phi}{\phi_m}\right)^2 ,\\
I(\phi) &= \left\{ 
\begin{array}{rl}
\left(\phi_m/\phi-1\right)^2 & \textrm{ if }\phi <\phi_m \\
0 & \textrm{ if }\phi \geq \phi_m,
\end{array} \right.
\end{align}
where $\phi_m= 0.585$, $\mu_1=0.3$ and $\delta=0.158$; these are plotted in Figure \ref{rheoplots}. We might have used a different rheology, but this model shows good agreement with the data of Boyer et al. \cite{boyer} and Dagois-Bohy et al. \cite{dagois} for $0.4<\phi<\phi_m$. Furthermore, owing to its linear extension beyond $\phi_m$, $\mu$ is a simple monotonic function, meaning we can invert it easily to find $\phi$. In other models $\phi(\mu)$ is constant for $\mu<\mu(\phi_m)$; this means that $\phi_m$ is the maximum volume fraction, regardless of how small shear stresses in the jammed slurry become. An important observation is that $\mu=0$ implies $\phi=\phi_m+\delta\mu_1\approx0.63\approx \phi_{rcp}$. Here $\phi_{rcp}$ is the random close packing limit, the maximal observed volume fraction due to random packing. This reflects the fact that, for a given confining stress, as the shear stress tends to zero, the particles pack to this maximal density. 

This rheology uses a continuum model that requires particles to be small compared to the size of the fracture. This is in order to well-define the proppant volume concentration, $\phi$. In our model the relevant ratio is that of the particle diameter to the typical crack width, the smallest cavity length scale. In \cite{lecgara}, good results are obtained using the same rheological model, with this ratio taking values as large as $1/10$. However, as the ratio approaches unity we have to consider non-local effects, such as proppant bridging across the crack width. This is particularly important near the fracture tip, where $w$ approaches zero. These effects will be discussed in greater detail in Section \ref{constgsec}, once we have formed a model of the evolving fracture. We must also be cautious applying these rheological models to dilute flows, since they are fitted to experimental data from the dense regime, where $\phi>0.4$. This difficulty is somewhat inevitable, since the determination of $I$ and $\mu$ requires measurement of the particle normal stress, or particle pressure, which becomes very small in the dilute regime.

\subsection{Fluid Slip}
We define $\mathbf{u}$ as the slurry velocity, $\mathbf{v}$ as the particle velocity and $\mathbf{q}=\mathbf{u}-\mathbf{v}$ as the slip velocity. We then employ the slip relation 
\begin{align}
\mathbf{q}&=\frac{a^2\kappa(\phi)}{\eta_f}\nabla\cdot\sigma^f, \label{slip}\\
\kappa(\phi)&=\frac{2(1-\phi)^{5.1}}{9\phi},
\end{align}
where $a$ is the particle radius and $\sigma_f$ is the fluid stress tensor. Since fluid and particle shear rates are often similar, we ignore fluid shear stresses and take $\sigma^f=-pI$; this is typical in the analysis of porous media flow. This simplifies (\ref{slip}) to Darcy's law. However, the effect of fluid shear stress is taken into account in the frictional rheology, where it is included as part of the solid shear stress. $\kappa$ is a normalised form of the permeability of the solid particles; we use the function suggested by Garside \& Al-Dibouni \cite{garside}, which is based on the phenomenology first described by Richardson \& Zaki \cite{zaki}. This choice of permeability function shows excellent agreement with the experimental results of Bacri et al. \cite{bacri}.


\subsection{Conservation Equations}
We consider the effective Reynolds number,
\begin{align}
\textrm{Re}_{\textrm{eff}} = \frac{\rho u_r w^2}{\eta_f R} \label{Re},
\end{align}
to be negligible. We also neglect the effect of gravity, since we are mainly concerned with small or neutrally buoyant proppants, which settle slowly. Hence, our momentum balance becomes
\begin{align}
\nabla \cdot \sigma =0,
\end{align}
where $\sigma=\sigma^s+\sigma^f$ is the mixture stress tensor, composed of the particle and fluid stresses respectively. We also note that, subtracting the hydrostatic pressure term, we write $\sigma = \tau -p I$. Since we assumed $\sigma^f=-pI$ in deriving the fluid slip equation, we deduce $\sigma_s=\tau$. This is a notational quirk arising from the frictional rheology because $\tau$ \textit{does} include shear stress originating from the viscous carrier fluid. Herein we will refer to $\sigma^s_{zz}$ and $\tau_{rz}$, since the former generally arises from the proppants and the latter stems from both the proppants and the carrier fluid. The assumption of axisymmetry gives
\begin{align}
\frac1r \frac{\partial(r \tau_{rr})}{\partial r} +\frac{\partial \tau_{rz}}{\partial z}-\frac{\partial p}{\partial r}&=0, &
\frac1r \frac{\partial(r \tau_{rz})}{\partial r} +\frac{\partial \sigma^s_{zz}}{\partial z}-\frac{\partial p}{\partial z}&=0 .
\end{align}
We also have the continuity equations
\begin{align}
\nabla \cdot (\mathbf{v}+\mathbf{q})&=0, &
\frac{\partial \phi}{\partial t}+\nabla \cdot (\phi \mathbf{v})&=0.
\end{align}
The first of these can be integrated over the fracture volume to give 
$
Qt=4\pi \int _0^R rw(r,t)dr.
$
Here, $Q$ is the rate at which the slurry is pumped into the crack, which we will assume is constant. We will also assume that the proppants are injected at a constant rate, meaning the average concentration at the wellbore is constant.


\section{Injection: Scalings}
\label{scalingsection}
To help implement the assumptions of a lubrication model, where the crack width is far smaller than the crack radius, we introduce the scaled coordinates,
\begin{align*}
T&= T(t),   &   r&= L(t) \Gamma(T) \xi, & z&= \epsilon(t) L(t) \eta.
\end{align*}
Here $T(t)$ is the internal time scale, a monotonic function to be specified later; $\epsilon(t)$ is a small number; and $\Gamma(T)$ is the crack radius, measured in the scaled coordinates, so $\xi=1$ implies $r=R$. We multiply the variables accordingly,
\begin{align*}
w(r,t) &\to \epsilon L w(\xi,T), &
p(r,z,t)&\to \epsilon E' p(\xi,\eta, T), &
R(t)&\to L \Gamma(T), 
\end{align*} 
\begin{align*}
v_z(r,z,t) &\to -\dot \epsilon L v_z(\xi, \eta, T), &
v_r(r,z,t) &\to \frac{-\dot \epsilon L}{\epsilon}v_r(\xi,\eta,T), \\ 
q_r(r,z,t)&\to \frac{\epsilon}{L}\frac{a^2E'}{\eta_f\Gamma}q_r(\xi,\eta,T),&
q_z(r,z,t)&\to \frac{1}{L}\frac{a^2E'}{\eta_f}q_z(\xi,\eta,T), \\
 \tau(r,z,t)&\to -\frac{\dot \epsilon}{\epsilon^2}\eta_f \tau(\xi,\eta,T), &
\sigma^s(r,z,t) &\to -\frac{\dot\epsilon}{\epsilon^2}\eta_f\sigma^s(\xi,\eta,T).
\end{align*}
The appearance of minus signs reflects the fact that $\epsilon$, the ratio of the characteristic radius to the characteristic width of the fracture, is decreasing. We also assume the scaling is suitable so that all the scaled variables are $\mathcal{O}(1)$. Herein, we will use $(\dot{  })$ for derivatives with respect to $t$ and $(')$ for those with respect to $T$. 

In the new, rescaled coordinates the equations describing the frictional rheology become 
$I(\phi)=\dot{\gamma}/\sigma_n^s $ and 
$\mu(\phi)=\tau/\sigma_n^s $.
The slip equation becomes
$\mathbf{q}=-\kappa(\phi)\nabla p,$
where $\nabla$ is now with respect to $(\xi,\eta)$. The integral equations become
\begin{align}
w(\xi,T)&=\frac{4\Gamma}{\pi} \int _{\xi}^{1} \frac{y}{\sqrt{ y^2-\xi^2}} \int_0 ^1 \frac{x p(x y , T)}{\sqrt{1-x^2}} dxdy,   &
\aleph \equiv \frac{K_{Ic}}{\epsilon E'\sqrt{L}}&=2 \sqrt{\frac{\Gamma}{\pi}} \int_0^1 \frac{p(\xi,T)\xi}{\sqrt{1-\xi^2}}d\xi.
\end{align}


The momentum equations are
\begin{align}
\frac{\epsilon}{\Gamma \xi} \frac{\partial (\xi \tau_{rr})}{\partial \xi}  +\frac{\partial \tau_{rz}}{\partial \eta}+\frac{\epsilon^3 E' t}{\eta_f } \frac{\epsilon}{\dot \epsilon t \Gamma} \frac{\partial p}{\partial \xi} &=0, &
\frac{\epsilon^2}{\Gamma\xi}\pdv{(\xi\tau_{rz})}{\xi}+\epsilon \pdv{\sigma^s_{zz}}{\eta}+\frac{\epsilon}{\dot\epsilon t}\frac{\epsilon^3 E' t}{\eta_f } \pdv{p}{\eta}&=0.
\end{align}
Since we expect the radial pressure gradient to be comparable to the shear stress, $\tau_{rz}$, we choose $\epsilon$ so that the dimensionless quantity $\epsilon^3 E' t/\eta_f=1$.
Finally, the global volume conservation equation then becomes
$Qt/(\epsilon L^3)=4\pi \Gamma^2 \int _0^1 \xi w(\xi,T)d\xi,$
so in a similar manner we choose the dimensionless quantity $Qt/\epsilon L^3=1.$
These choices mean
\begin{align}
\epsilon(t) &=(\eta_f/E')^\frac13 t^{-1/3}, &
L(t)&=(E'Q^3/\eta_f)^\frac19 t^{4/9}.
\end{align}
We will repeatedly use the relations $\dot \epsilon t/\epsilon=-1/3$ and $\dot L t/L=4/9$. Using this choice of $\epsilon$ we note that, before scaling, $\sigma^s/p=\mathcal{O}(\epsilon)$; this validates the assumption that particle pressure is negligible compared to hydrostatic pressure at the crack walls. Also, by the scaled momentum equations,
\begin{align}
\frac{\partial\tau_{rz}}{\partial \eta}&=\frac{3}{\Gamma}\frac{\partial p}{\partial \xi}+\mathcal{O}(\epsilon) , &
\frac{\partial p}{\partial \eta}&= \frac{\epsilon}{3} \frac{\partial \sigma^s_{zz}}{\partial \eta} +\mathcal{O}(\epsilon^2), \label{etadivp}
\end{align}
the second of which verifies the assumption that $p$ has spatial dependence in the radial direction only. Because of the $\eta=0$ reflectional symmetry, we note that $\tau_{rz}(\xi,0)=0$. So, ignoring $\mathcal{O}(\epsilon)$ terms and integrating (\ref{etadivp}.1), we see that
\begin{align}
\tau_{rz}=\frac{3\eta}{\Gamma}\frac{\partial p}{\partial \xi},
\end{align}
and, using the scaled equations from the frictional rheology,
\begin{align}
\sigma_{zz}^s&=\frac{3|\eta|}{\Gamma} \frac{1}{\mu(\phi)} \frac{\partial p}{\partial \xi}, \label{normalstress} &
\frac{\partial v_r}{\partial \eta} &=\frac{3\eta}{\Gamma} \frac{I(\phi)}{\mu(\phi)} \frac{\partial p}{\partial \xi}.
\end{align}
Then, using the condition $v_r(\xi,\pm w)=0$, we deduce that
\begin{align}
v_r(\xi,\eta)=-\frac{3}{\Gamma}\pdv{p}{\xi}\int_\eta^w \frac{I(\phi)\eta}{\mu(\phi)} d\eta. \label{vrexpression}
\end{align}


\section{Injection: Time Regimes}
\label{timeregimes}

In this choice of scaling, the slurry conservation equation becomes
\begin{align}
\frac{1}{3\Gamma \xi}\frac{\partial(\xi v_r)}{\partial \xi}+\frac{1}{3}\frac{\partial v_z}{\partial \eta}+\left(\frac{a}{L\Gamma}\right)^2 \frac{1}{\epsilon^2\xi}\frac{\partial(\xi q_r)}{\partial \xi}+\left(\frac{a}{L}\right)^2\frac{1}{\epsilon^4}\frac{\partial q_z}{\partial \eta}=0.
\end{align}
Combining this with the scaled slip equation, noting (\ref{etadivp}), we obtain
\begin{align}
\frac{1}{3\Gamma \xi}\frac{\partial(\xi v_r)}{\partial \xi}+\frac{1}{3}\frac{\partial v_z}{\partial \eta}-\frac{\epsilon \lambda }{\Gamma^2 \xi}\pdv{}{\xi}\left[\xi\kappa(\phi)\pdv{p}{\xi}\right]-\frac{\lambda}{3}\pdv{}{\eta}\left[\kappa(\phi)\pdv{\sigma^s_{zz}}{\eta}\right]=0. \label{slurrycont}
\end{align}
Here $\lambda=a^2/(L^2\epsilon^3)$ is a constant; we will later identify it as the ratio of the fracture length scale to the development length scale, over which we expect proppant flow to stabilise.

According to Shiozawa \& McClure \cite{shio}, Chen Zhixi et al. \cite{chen} and Liang et al. \cite{proppants}, we utilise the following constants, relevant to hydraulic fracturing, as given in Table \ref{values}.
%
 %
\begin{table}
{\renewcommand{\arraystretch}{1.15}
\begin{tabular}{|L|L|}
\hline
\textrm{Constant} & \textrm{Typical Value} \\
\hline
Q & 0.04\textrm{m}^3\textrm{ s}^{-1} \\
E' & 40\textrm{ GPa} \\
\eta_f & 0.01\textrm{ Pa s} \\
\rho_f & 1000 \textrm{ kg m}^{-3} \\
K_{Ic} & 0.5 \textrm{ MPa m}^{0.5} \\
a & 5\times 10^{-5}\textrm{m}\\
\hline
\end{tabular}}
\caption{Typical values of constants, given by Shiozawa \& McClure \cite{shio}, Chen Zhixi et al. \cite{chen} and Liang et al. \cite{proppants}.}
\label{values}
\end{table}
%
%
The choice of $a$ represents a typical diameter for the finer proppants commonly used at the initiation of fracturing \cite{proppants}. This gives us the following estimates
\begin{align*}
\epsilon &\approx 6\times10^{-5} \cdot t^{-1/3}, &
L &\approx 9 \times 10^0 \cdot t^{4/9}, \\
\textrm{Re}_\textrm{eff} &\approx 1\times10^{-2} \cdot t^{-7/9}, &
 \aleph &\approx 4\times10^{-2} \cdot t^{1/9}, \\
\lambda &\approx 1\times10^2 \cdot  t^{1/9}, &
a/(\epsilon L)& \approx 1\times 10^{-1} \cdot t^{-1/9} . 
\end{align*}

The value of $\textrm{Re}_\textrm{eff}$ is calculated using formula (\ref{Re}), substituting each term with its typical scaling.

Considering the same problem in the zero-proppant case, Detournay \& Savitski \cite{detsav} show that when $1.6\aleph <1$, the fracture evolution is well approximated by taking the dimensionless toughness $\aleph=0$. Also, the choice $T=\aleph$ is taken, reflecting the dependence of the scaled solution on this monotonically increasing parameter; assuming $\aleph$ is negligible it is possible to neglect any $T$ dependence. We will also use these assumptions, since toughness plays its greatest role near the fracture tip, where the crack is typically too narrow for proppants to interfere. Given our estimate for $\aleph$, this means we must take $t<1.5\times10^7$.

In general we will assume $t>250$, so we may ignore $\epsilon$ and $\textrm{Re}_\textrm{eff}$ terms. This also means $2a/(\epsilon L)<1/10$, so the fracture is typically more than 10 particles wide. Lecampion \& Garagash \cite{lecgara}, conclude that non-local phenomena such as proppant-bridging aren't important in such cases; however we can still expect to see these effects near the narrow crack tip. The significance of this behaviour will be discussed in greater detail in Section \ref{constgsec}.


We also note that $\lambda$ is large; so in an effort to remove time dependence from our equations, we may neglect the first three terms in the continuity equation (\ref{slurrycont}), 
\begin{align}
\pdv{}{\eta}\left[\kappa(\phi)\pdv{\sigma^s_{zz}}{\eta}\right]=0.
\end{align}

By the assumption of reflectional symmetry, the particle pressure gradient must vanish at $\eta=0$. Because $\kappa$ is generally non-zero, we deduce that the particle pressure is constant with $\eta$; and, by (\ref{normalstress}), so is $|\eta|/\mu(\phi)$. Hence,
\begin{align}
\phi (\xi , \eta)=\mu^{-1}\left(\mu_w(\xi)\frac{ |\eta|}{w(\xi)}\right), \label{fintroduction}
\end{align}
where $\mu_w$ is an undetermined function of $\xi$, which we recognise as the value of $\mu$ at the crack wall. Noting that $\mu$ is a decreasing function, we see that $\mu_w$ also describes the rate at which the concentration drops from the centre to the wall of the cavity. We also notice that, in accordance to Donstov et al. \cite{donstov}, we have plug flow in the centre of the channel, where concentrations are greater than $\phi_m$. Because the slurry flows away from the wellbore, the distribution of proppants, which is described by $\mu_w$, depends on the concentration of proppants in the injected mixture and how that changes with time. Hence, an important step in the determination of $\mu_w$ will be implementing the assumption that the average concentration at the wellbore is constant. This will be discussed in greater detail in Section \ref{constgsec}. 

It is interesting to note that \cite{lecgara} verifies a length scale of $\epsilon^3L^3/a^2$ for proppant flow in a channel, or pipe, to become fully established. This means the particle pressure gradient becomes negligible, and the cross fracture concentration profile becomes independent of the distance from the channel, or pipe, entrance. As a result, the constant $\lambda=a^2/(L^2\epsilon^3)$ can be interpreted as the ratio of the fracture length to the development length. Because this is large, an alternative route to (\ref{fintroduction}) would have been to assume the transverse particle pressure is constant, reflecting the full development of the flow.


\section{Injection: Governing Equation for fracture width}
\begin{figure}
\begin{subfigure}{0.49\textwidth}
\includegraphics[width=\textwidth]{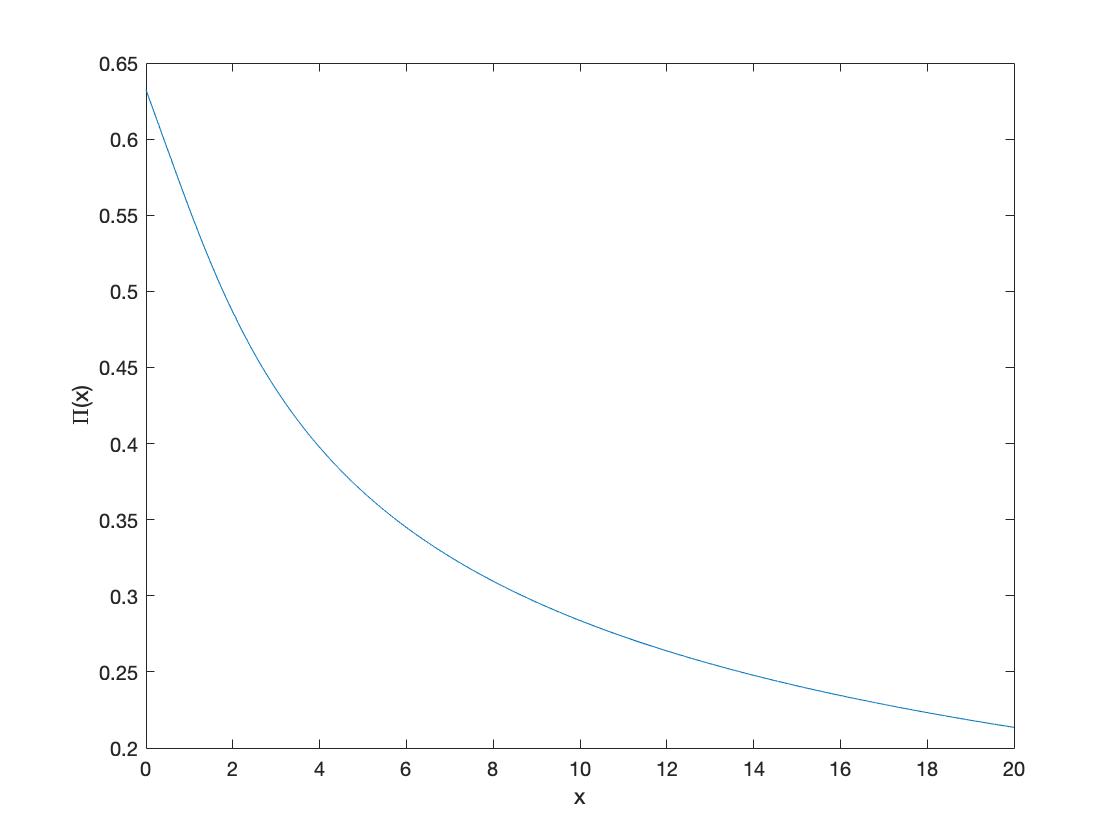}
\caption{$\Pi(x)$ as a function of $x$}
\end{subfigure}
\begin{subfigure}{0.49\textwidth}
\includegraphics[width=\textwidth]{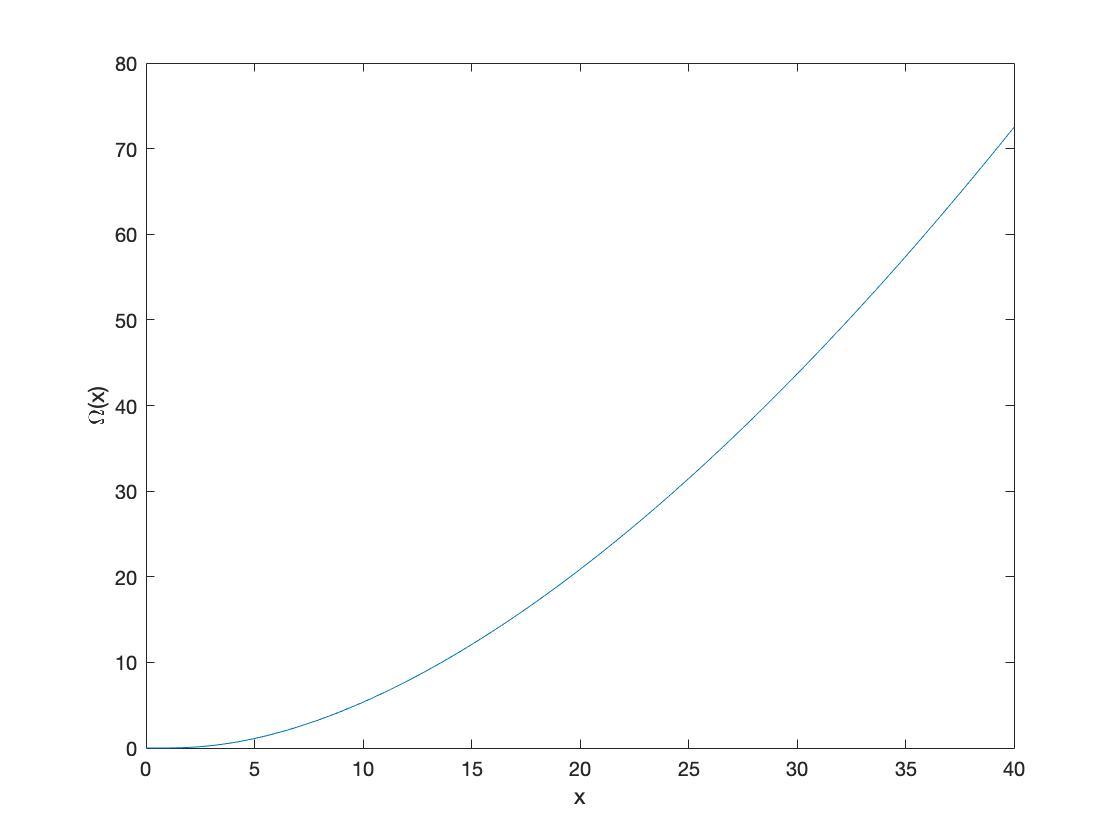}
\caption{$\Omega(x)$ as a function of $x$}
\end{subfigure}
\begin{subfigure}{0.49\textwidth}
\includegraphics[width=\textwidth]{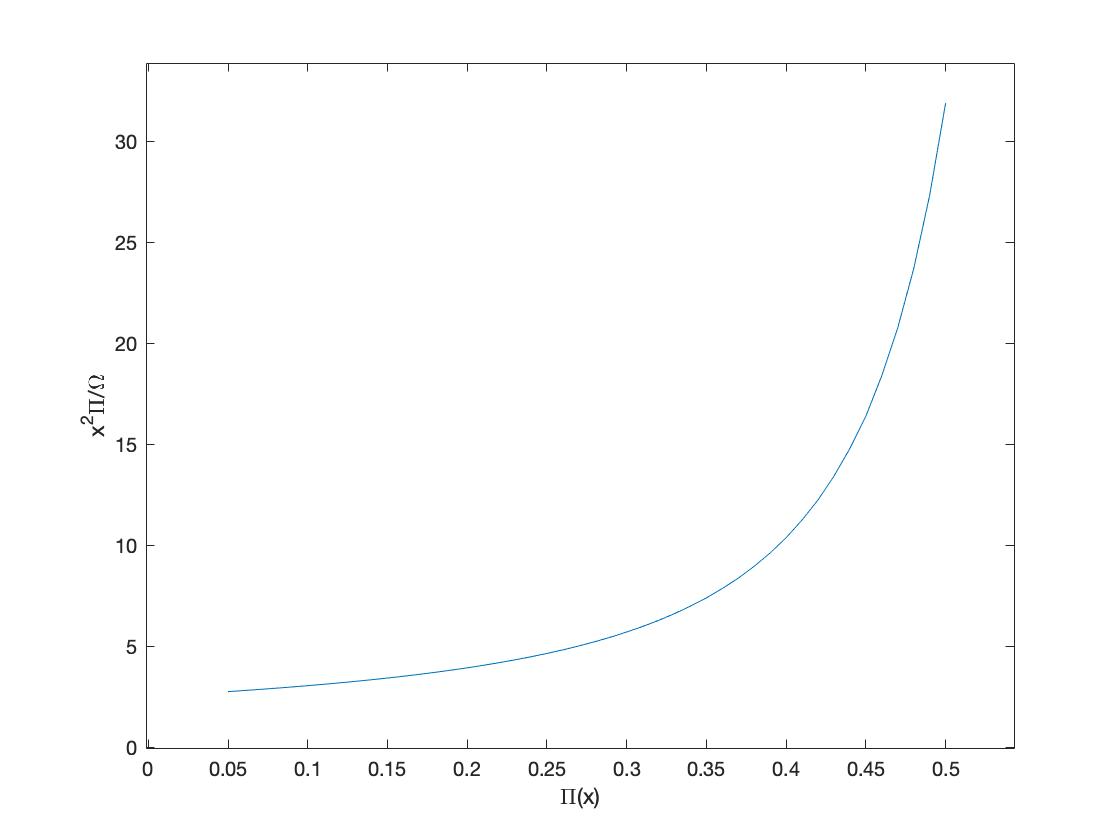}
\caption{$x^2 \Pi / \Omega$ as a function of $\Pi(x)$}
\label{omegapisub}
\end{subfigure}
\centering
\caption{Plots of the rheological functions $\Omega$, $\Pi$ and $x^2 \Pi / \Omega$.}
\label{omegapi}
\end{figure}

In scaled coordinates, the governing equation for the conservation of proppant mass becomes
\begin{align}
\frac{\xi \dot{L} t}{L}\frac{ \partial \phi}{\partial \xi} +\left[ \frac{ \dot{\epsilon} t}{\epsilon}+\frac{ \dot{L} t}{L} \right] \eta \frac{ \partial \phi}{\partial \eta} = -\frac{\dot{\epsilon}t}{\epsilon \Gamma\xi}\pdv{(\xi \phi v_r)}{\xi} -\frac{\dot{\epsilon}t}{\epsilon}\pdv{(\phi v_z)}{\eta}. \label{prebigwidth}
\end{align}
Then, implementing our choices of $\epsilon$ and $L$, we obtain
\begin{align}
\frac{4\xi}{3} \frac{ \partial \phi}{\partial \xi} +\frac\eta3 \frac{ \partial \phi}{\partial \eta} = \frac{1}{\Gamma\xi}\pdv{(\xi \phi v_r)}{\xi} +\pdv{(\phi v_z)}{\eta}.
\end{align}
Integrating from $-w$ to $w$ with respect to $\eta$, leaving details to Appendix A for brevity, we obtain
\begin{align}
4\xi \pdv{}{\xi}\left[w \Pi\circ \mu_w(\xi) \right]-w\Pi\circ \mu_w(\xi)=-\frac{9}{\Gamma^2 \xi} \pdv{}{\xi}\left[\frac{\xi w^3}{\mu_w(\xi)^2} \pdv{p}{\xi}\Omega\circ \mu_w(\xi)\right] \label{bigwidth}.
\end{align}
Here we have defined the rheological functions
\begin{align}
\Pi(x)&=\frac{1}{x} \int _0^x \mu^{-1}(u) du, &
\Omega(x)&=\frac{1}{x}\int_0^x [\Pi(u)I\circ \mu^{-1}(u)u]du,
\end{align}
which we plot in Figure \ref{omegapi}.

Multiplying by $\xi$ and integrating from $\rho$ to $1$, we obtain
\begin{align}
 \int _\rho ^1 \xi w \Pi \circ \mu_w (\xi) d\xi+\frac{4}{9} \rho^2 w \Pi \circ \mu_w(\rho)  =-\frac{\rho w^3}{\Gamma^2 \mu_w^2} \pdv{p}{\rho} \Omega \circ \mu_w (\rho), \label{bigwidth2}
\end{align}
which lends itself more easily to computation. Here we have taken $w^3 \partial p/\partial \xi \to 0$ as $\xi\to1$; this is physically motivated by the fact that this term is proportional to the radial flux, which vanishes at the crack tip. Moreover, Spence \& Sharp \cite{1985} show that, in the zero-proppant, zero-toughness regime, near the crack tip, $p\propto(1-\xi)^{-1/3}$ and $w\propto (1-\xi)^{2/3}$. 

In order to compare this equation to the zero-proppant case, we assume $\mu_w$ is independent of $\xi$ and take $\mu_w\to \infty$, to obtain
\begin{align}
\int_\rho^1 \xi w(\xi) d\xi +\frac49 \rho^2 w=-\frac{ \rho w^3}{\Gamma^2}\pdv{p}{\rho} \lim_{\mu_w\to\infty} \left[\frac{\Omega (\mu_w)}{\mu_w^2 \Pi(\mu_w)}\right]. \label{fluidlim}
\end{align}
From Figure \ref{omegapisub} we deduce the right hand limit is approximately $2/5$, which is confirmed exactly in Appendix B. Modelling the fluid as Newtonian, also leaving the details to Appendix B, we obtain the same equation, with a factor of $1/3$ instead. We conclude that the equations governing Newtonian flow are not the same as those in the zero-proppant slurry flow limit. This is clearly a limitation of our approach, which arises from using a dense-fitted rheology in the dilute regime. However, the fact that the equations share a nearly identical form is promising, as we expect the qualitative behaviour of slurry flow to be similar to that of Newtonian flow.


\section{Injection: Numerical Solution}
\label{numsol}
We implement the numerical method first used by Spence \& Sharp \cite{1985}, with the adaptions of Detournay \& Savitski \cite{detsav}, to solve the equations we have derived so far. It will be useful to introduce $h(\xi)=w(\xi)/\Gamma$. The lubrication equation derived above, the elasticity equations and the global volume conservation equation become
\begin{align}
 \int_\rho^1 (\xi h \Pi\circ \mu_w) d\xi &+ \frac49 \rho^2 h\Pi \circ \mu_w   =-\rho h^3\pdv{p}{\rho}\frac{\Omega \circ \mu_w}{\mu_w^2}, \label{pigquation} \\
h(\xi)&=\frac{4}{\pi} \int _{\xi}^{1} \frac{y}{\sqrt{ y^2-\xi^2}} \int_0 ^1 \frac{x p(x y)}{\sqrt{1-x^2}} dxdy, \label{underpressure}  \\
0&=\int_0^1 \frac{p(\xi)\xi}{\sqrt{1-\xi^2}}d\xi, \\
1&=4\pi \Gamma^3 \int _0^1 (\xi h)d\xi. \label{pigquation2}
\end{align}
These equations alone do not give unique solutions for $\{p,h,\mu_w\}$, so we will prescribe $\mu_w$ as part of the problem data. This allows us to uniquely determine a solution for $\{p,h\}$. We seek series approximations of the form
\begin{align}
p(\xi)&=\sum _{i=-1}^{N-1} A_i p_i(\xi), &
h(\xi)&=\sum_{i=-1}^{N} B_i h_i(\xi), 
\end{align}
where we define
\begin{align}
p_i(\xi)&=\left\{ 
\begin{array}{ll}
-\ln\xi +\ln2-1&(i=-1) \\
&\\
(1-\xi)^{-1/3} J_i(\frac43,2,\xi)+\omega_i & (i\geq0) 
\end{array} \right\},
&h_i(\xi)=\left\{ 
\begin{array}{ll}
\frac4\pi \left[(1-\xi^2)^{1/2}-\xi \cos^{-1}(\xi) \right] &(i=-1) \\
&\\ \nonumber \\
(1-\xi)^{2/3} J_i(\frac{10}3,2,\xi) & (i\geq0) 
\end{array} \right\}.
\end{align}
Here the $i=-1$ terms are used to account for the logarithmic singularity in pressure at the inlet, expected as a result of the point source injection; the other terms allow for a general solution of (\ref{underpressure}). Importantly, we note that the $p_i$ terms have a $(1-\xi)^{-1/3} $ singularity near the crack tip and the $h_i$ terms are proportional to $(1-\xi)^{2/3} $ (for $i\geq0$). This deliberately matches the asymptotic calculations from Spence \& Sharp \cite{1985}, which arise from the assumptions of zero-lag and zero-toughness in an expanding hydraulic fracture. This allows the numerical method to converge accurately with few terms. The $J_i(p,q,\xi)$ are Jacobi Polynomials of order $i$ defined on the interval $[0,1]$, in the sense defined by Abramowitz \& Stegun \cite{abram}, normalised to satisfy the orthonormality condition,
\begin{align}
\int_0^1 (1-\xi)^{p-q}\xi^{q-1}J_i(p,q,\xi)J_j(p,q,\xi)d\xi=\delta_{ij}. 
\end{align}
This means that the $h_{i}$ ($i\geq 0$) are orthonormal with respect to an inner product weighted by $\xi$. The $\omega_i$ are simply constants to ensure each of the $p_i$ obey the zero-toughness equation; adding these constants means that the $p_i$ lose their orthonormality properties, however this doesn't affect the solution finding process.

Because of its linearity, these series approximations reduce (\ref{underpressure}) to a linear equation,
\begin{align}
B_i=\sum_{j=-1}^{N-1}P_{ij}A_j.
\end{align}
Here $(P)_{ij}$ is an $(N+2)\times(N+1)$ matrix whose entries we only have to calculate once by using the orthogonality relation given above, along with the fact that $\{p_{-1},\theta_{-1}\}$ are a solution pair to (\ref{underpressure}). The entries of $M$, which can be found in \cite{detsav}, are listed in Appendix C for $N=4$. The subtleties of calculating elements of $P_{ij}$, in the face of strong singular behaviour, are important and described in depth in \cite{detsav}. Finally, using the values of $B_i$ given above, we assign a cost to each choice of $A$ given by 
\begin{align}
\Delta(A)=\sum _{\xi \in \{0,1/M,...,1\} } \left( \frac{\textrm{RHS}(\xi ; A)}{\textrm{LHS}(\xi ; A)}-1\right)^2.
\end{align}
This is calculated by considering the discrepancies between the left and right hand sides of (\ref{pigquation}), calculated at M+1 equally spaced control points. We then minimise $\Delta$ with respect to $A$ using the Nelder-Mead Simplex method \cite{nelder}.


\section{Injection: Solutions for a constant \texorpdfstring{$\mu_w$}{muw}}
\label{constgsec}

\begin{figure}
\includegraphics[width=\textwidth,height=0.4\textwidth]{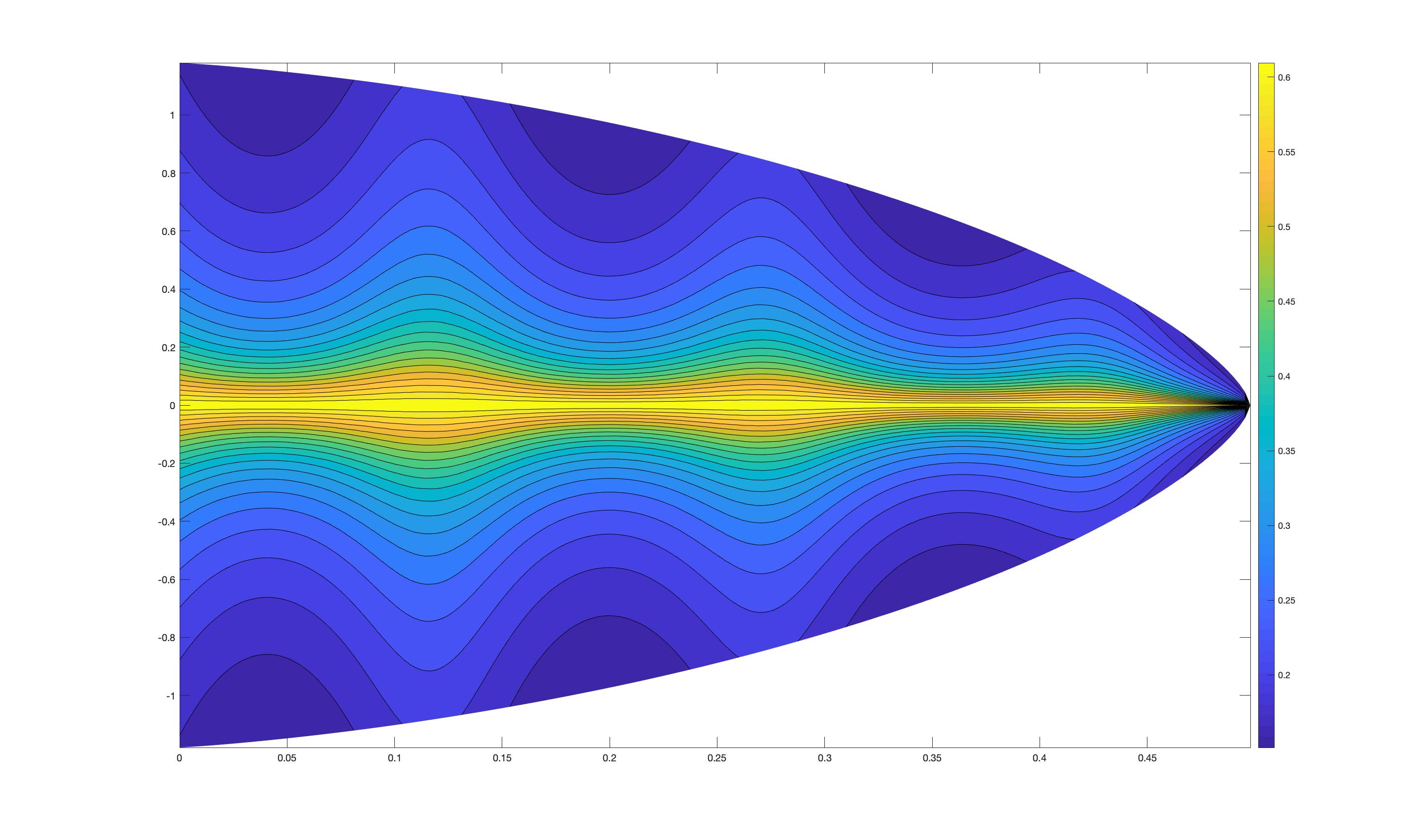}
\centering
\caption{Plot of cavity width profile and proppant distribution in the case where $\mu_w$ is sinusoidal. Here $N=8$ is used.}
\label{sing}
\end{figure}

For most monotonic choices of $\mu_w$, the numerical method above shows good convergence. We see that the coefficients $A_i$ and $B_i$ drop off quickly with $i$, and the final value of $\Delta$ tends to zero rapidly as we increase $N$. If $\mu_w$ is a more complicated function, like in the case of Figure \ref{sing}, we may need to use a larger value of $N$, but good convergence is still possible. 

This leads us to consider which choices of $\mu_w$ are most likely to appear in reality. We note that by (\ref{fintroduction}), 
\begin{align}
\Pi\circ \mu_w(\xi)=\frac{1}{2w}\int_{-w}^w \phi(\xi, \eta)d\eta, \label{pimeaning}
\end{align}
so we may view $\Pi\circ \mu_w(\xi)$ as the average proppant concentration at a given value of $\xi$. Since $\Pi\circ \mu_w$ is independent of time, we automatically satisfy the condition that the injection rates of the proppants and the fluid are constant. However this condition also means that the average concentration at the wellbore, $\Pi \circ \mu_w(0)$, must equal the average concentration taken by integrating over the entire crack volume. For a monotonic choice of $\mu_w$ this implies that $\mu_w$ must be independent of $\xi$. Herein we will make the assumption that $\mu_w$ is a constant and, as a result, so is $\Pi=\Pi(\mu_w)$. This is a natural assumption: at early times we don't expect significant concentration differences along the crack because radial length scales are small.

A great advantage of a constant $\Pi$ is that we can define an `effective viscosity', which we can absorb into our scaled variables the same way as we did with fluid viscosity. Under the assumption that $\mu_w$ is constant, (\ref{pigquation}) becomes
\begin{align}
\int_\rho^1 \xi h(\xi) d\xi +\frac 49 \rho^2 h=- \frac{\rho h^3}{\eta_e}\pdv{p}{\rho},  \label{mueffeq}
\end{align}
where $\eta_e = \mu_w^2\Pi/\Omega$
is what we call the effective viscosity. It is plotted in Figure \ref{omegapisub}, and is best thought of as a function of the average concentration, $\Pi$. Making the transformations
\begin{align}
h&=\eta_e^{1/3} \tilde h, & p&=\eta_e^{1/3}\tilde p, & \Gamma&=\eta_e^{-1/9}\tilde \Gamma,
\label{effviscrels}
\end{align}
our governing equations become
\begin{equation}
\begin{aligned}
\int_\rho^1 \xi \tilde h d\xi &+ \frac49 \rho^2 \tilde h   =-\rho \tilde h^3\pdv{p}{\rho},&
\tilde h(\xi)&=\frac{4}{\pi} \int _{\xi}^{1} \frac{y}{\sqrt{ y^2-\xi^2}} \int_0 ^1 \frac{x \tilde p(x y)}{\sqrt{1-x^2}} dxdy,   \\
0&=\int_0^1 \frac{\tilde p(\xi)\xi}{\sqrt{1-\xi^2}}d\xi, &
1&=4\pi \tilde \Gamma^3 \int _0^1 (\xi \tilde h)d\xi.
\end{aligned} \label{absorbsystem}
\end{equation}
We will solve them using the numerical method described before, except with (\ref{absorbsystem}) in the place of (\ref{pigquation}-\ref{pigquation2}).

\begin{figure}
\begin{subfigure}{0.49\textwidth}
\includegraphics[width=\textwidth,height=0.6\textwidth]{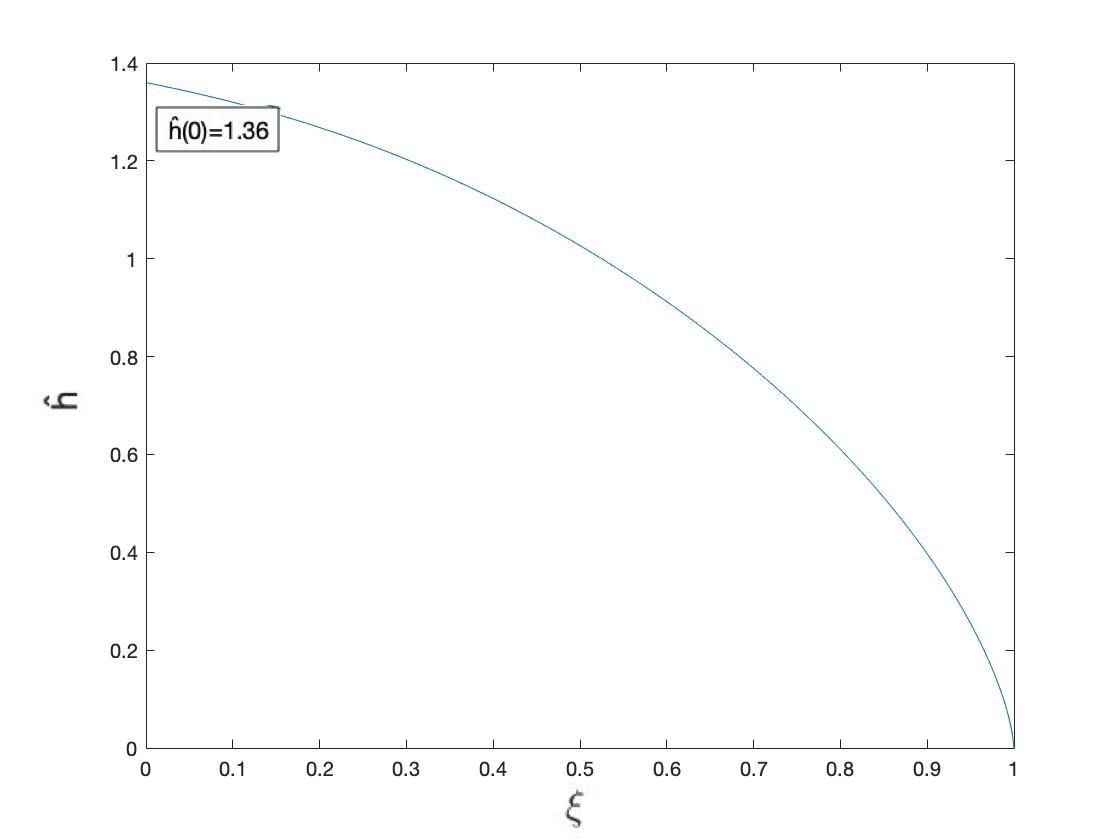}
\end{subfigure}
\begin{subfigure}{0.49\textwidth}
\includegraphics[width=\textwidth,height=0.6\textwidth]{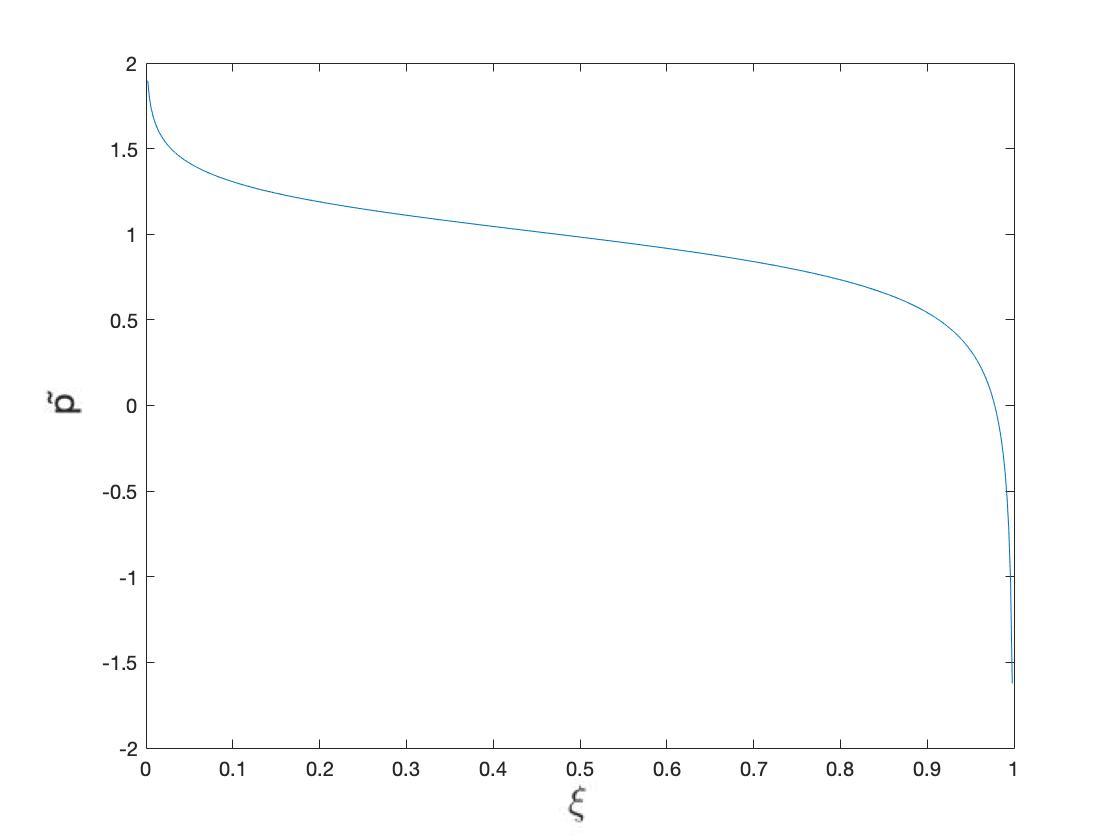}
\end{subfigure}
\caption{$(\xi,\eta)$ plots of $\tilde h$ and $\tilde p$, the scaled width and pressure solutions to the absorbed effective viscosity system.}
\label{tildes}
\end{figure}

Figure \ref{tildes} plots $\tilde h$ and $\tilde p$, calculated using $N=4$ and $M+1=501$ control points. Promisingly, we note that $\tilde h>0$ and $p$ shows the expected asymptotic behaviour. The value $\tilde h(0)=1.36$ will be important in later discussion. The first column of table \ref{alpha1tab} shows the coefficients $A_i$ and $B_i$, as well as the calculated value of $\tilde \Gamma =0.598$. Significantly, we see that $A_i$ and $B_i$ decrease rapidly with $i$, suggesting that a solution with higher order terms is unnecessary. This is supported by the small value of $\Delta \approx 5\times 10^{-5}$, with evenly spread contributions from control points along the radius of the crack. This suggests that we have found a genuine solution, and that the tip asymptotics are indeed suitable.

We now focus on finding numerical solutions for different concentrations in order to consider features such as the velocity profile and proppant distribution within the cavity. We consider the case of four different values of the average concentration, $\Pi$. These are given in table \ref{Pivals}, along with the corresponding values of $\mu_w$ and $\eta_e$.

\begin{table}
{\renewcommand{\arraystretch}{1.15}
\begin{tabular}{|c|c|c|}
\hline
$\Pi$ & $\mu_w$ & $\eta_e$\\
\hline
0.05	& 487.3 & 2.74\\
0.20 & 23.35 & 3.92\\
0.40	& 3.93 & 10.37\\
0.55 & 1.06 & 96.60\\
\hline
\end{tabular}}
\caption{Test values of $\Pi$, $\mu_w$ and $\eta_e$.}
\label{Pivals}
\end{table}

The latter columns of table \ref{alpha1tab} show the values of $A$, $B$ and $\Gamma$ calculated using the exact method suggested in Section \ref{numsol}. Again we use $M+1=501$ control points and $N=4$. Happily, the same values are observed by using the values of $A$, $B$ and $\Gamma$ listed in the first column, calculated after absorbing the effective viscosity, and using the relations (\ref{effviscrels}) to return to the concentration-specific values. We calculate the same value of $\Delta \approx 5\times 10^{-5}$ each time; this is to be expected as the equations are equivalent once the solutions have been scaled.

\begin{figure}
\begin{subfigure}{0.49\textwidth}
\includegraphics[width=\textwidth,height=0.6\textwidth]{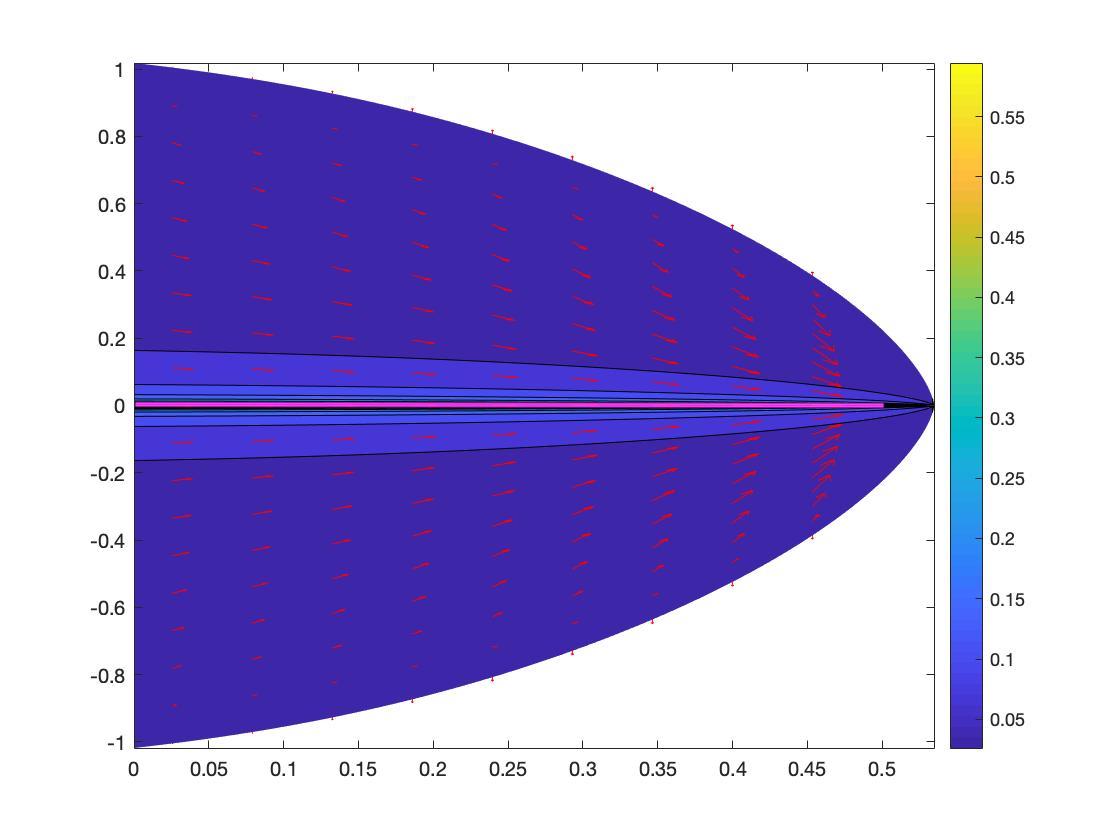}
\caption{$\Pi=0.05$}
\label{alpha1first}
\end{subfigure}
\begin{subfigure}{0.49\textwidth}
\includegraphics[width=\textwidth,height=0.6\textwidth]{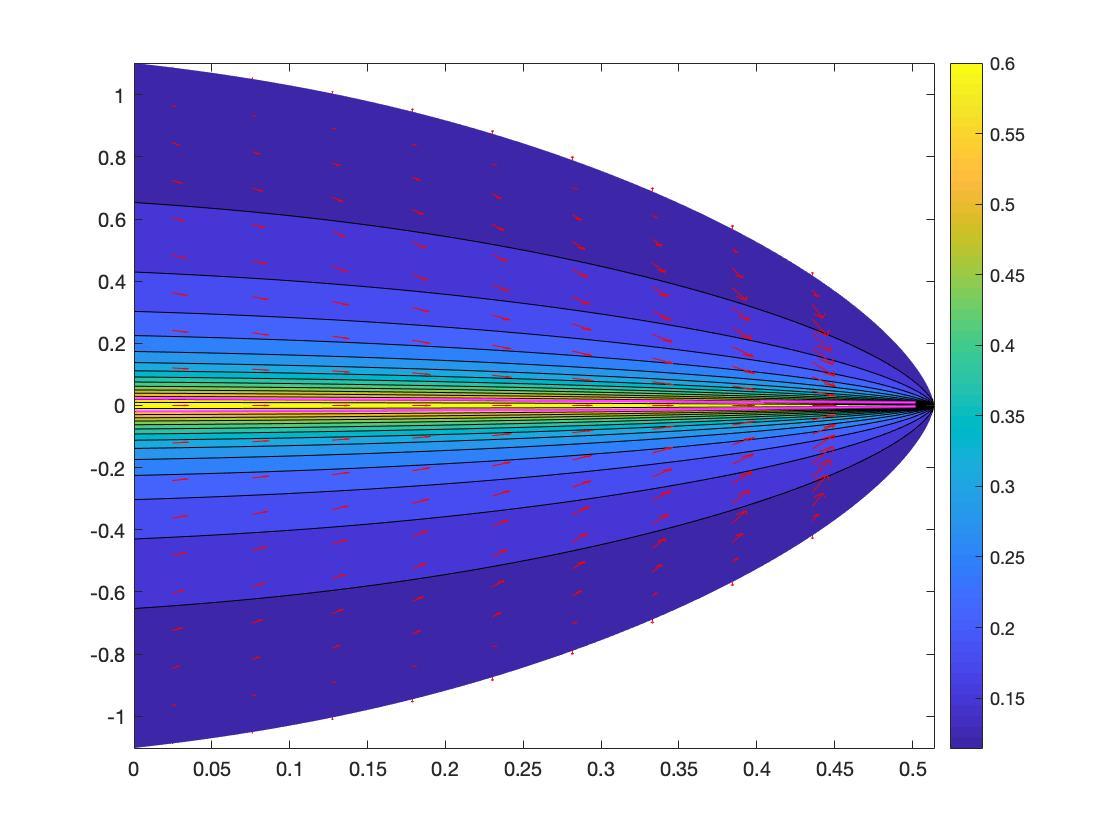}
\caption{$\Pi=0.20$}
\end{subfigure}
\begin{subfigure}{0.49\textwidth}
\includegraphics[width=\textwidth,height=0.6\textwidth]{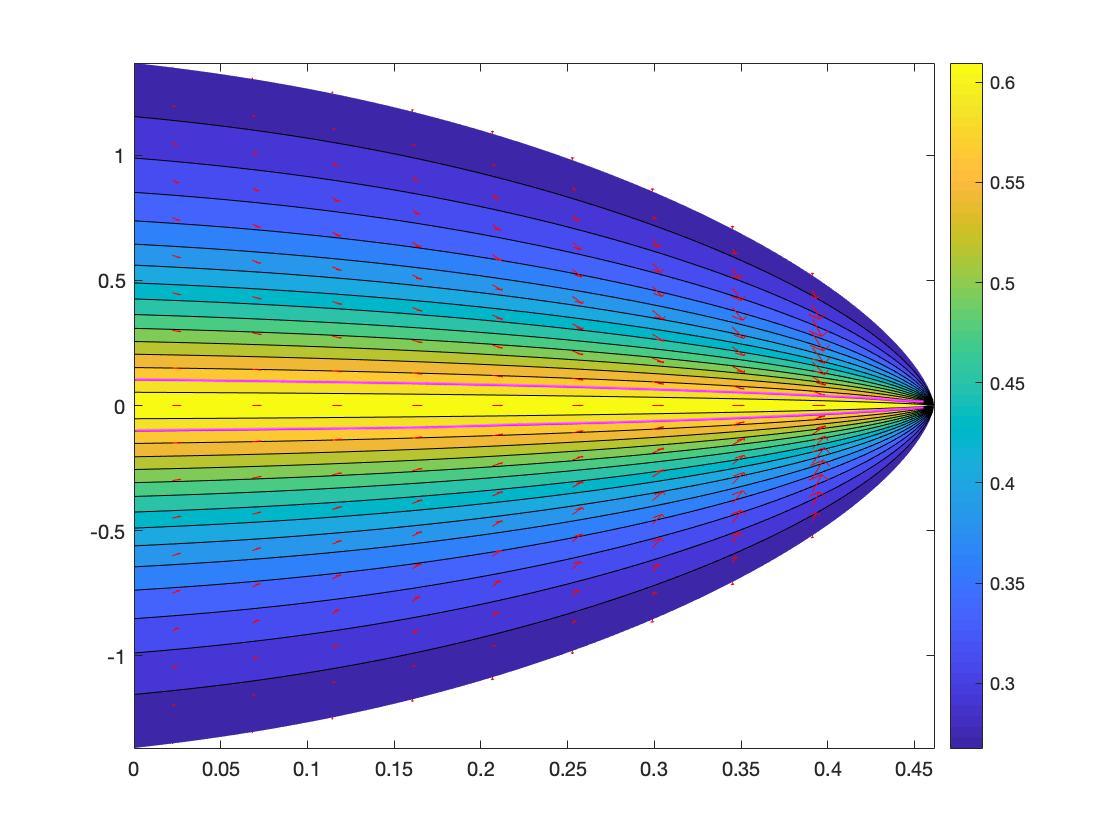}
\caption{$\Pi=0.40$}
\end{subfigure}
\begin{subfigure}{0.49\textwidth}
\includegraphics[width=\textwidth,height=0.6\textwidth]{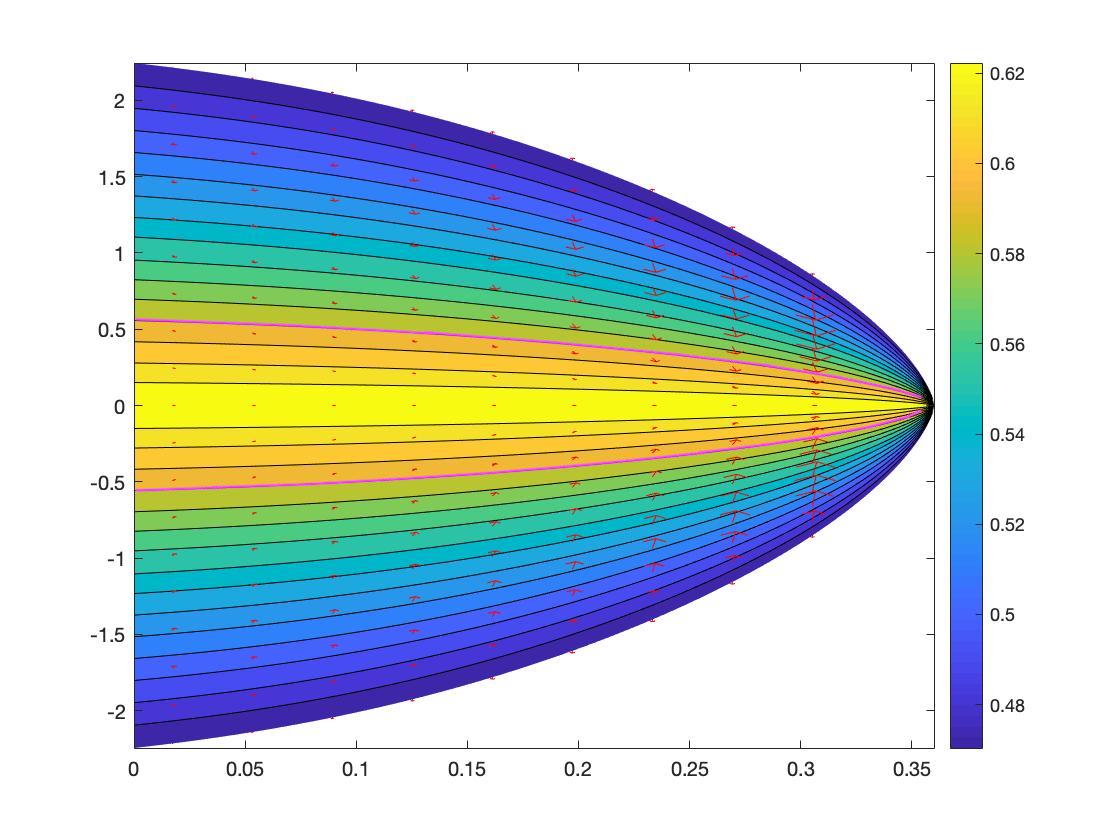}
\caption{$\Pi=0.55$}
\label{alpha1last}
\end{subfigure}
\caption{Concentration-specific $(\Gamma \xi, \eta)$ plots of developing fractures with total solid volume fraction, $\Pi$, taking the values $0.05$, $0.20$, $0.40$ and $0.55$. These are presented with filled contours displaying proppant concentration; arrows showing $\xi$-scaled velocity; and magenta contours indicating the transition into plug flow at the centre of each cavity.}
\label{alpha}
\end{figure}

\begin{table}
\begin{tabular}{|ccc||c||c|c|c|c|}
\hline
 &&&&&&&\\
 & $\Pi$ && Absorbed  & 0.05 & 0.20 & 0.40 & 0.55 \\
 &&&&&&&\\
 \hline
 &&&&&&&\\
 & $A_{-1}$ & &0.14786  & 0.20710 & 0.23326 & 0.32238 & 0.67830 \\
 & $A_{0}$&  &0.53529 & 0.74974 & 0.84444 & 1.16709 & 2.45559 \\
 & $A_{1}$&  &0.01929 & 0.02702 & 0.03043 & 0.04206 & 0.08849 \\
 & $A_{2}$&  &0.00402 & 0.00563 & 0.00634 & 0.00877 & 0.01844 \\
 & $A_{3}$&  &0.00035 & 0.00049 & 0.00055 & 0.00076 & 0.00159 \\
 &&&&&&& \\
 \hline
 &&&&&&& \\
 & $B_{-1}$ &  &0.14786 & 0.20710 & 0.23326 & 0.32238 & 0.67830 \\
 & $B_{0}$&  &0.53805 & 0.75361 & 0.84879 & 1.17311 & 2.46825 \\
 & $B_{1}$&  &0.05435 & 0.07612 & 0.08573 & 0.11849 & 0.24931 \\
 & $B_{2}$&  &0.00012 & 0.00016 & 0.00019 & 0.00026 & 0.00054 \\
 & $B_{3}$&  &0.00081 & 0.00114 & 0.00128 & 0.00177 & 0.00373 \\
 & $B_{4}$&  &0.00029 & 0.00041 & 0.00046 & 0.00064 & 0.00134 \\
 &&&&&&&\\
 \hline
 &&&&&&& \\
 & $\Gamma$ &  &0.59812 & 0.534579 & 0.513799 & 0.461261 & 0.359968 \\
  &&&&&&& \\
  \hline
 \end{tabular}
\caption{Values of $A_i$, $B_i$ and $\Gamma$ obtained using (\ref{absorbsystem}) with effective viscosity absorbed into the scaling and (\ref{pigquation}-\ref{pigquation2}) with $\Pi \in \{0.05,0.20,0.40,0.55\}$. We use $M=500$ and $N=4$ throughout.}
\label{alpha1tab}
\end{table}
Figure \ref{alpha} shows the distribution of proppants within the fracture for each value of $\Pi$. They are overlaid with an arrow plot of the proppant velocity profile, $\mathbf{v}$, scaled by $\xi$ to show the equivalent two-dimensional flux. The calculation of $\mathbf{v}$ is omitted since it is lengthy and similar to the derivation of (\ref{bigwidth}) in Appendix A. As $\Pi$ increases we see a growing disk of plug flow where $\phi>\phi_m$, marked with a magenta contour. We also see a tendency towards proppant velocity across the crack, rather than along it; this is because the shape of the crack becomes shorter and wider as the effective viscosity increases.


\begin{figure}
\begin{subfigure}{0.48\textwidth}
\includegraphics[width=\textwidth]{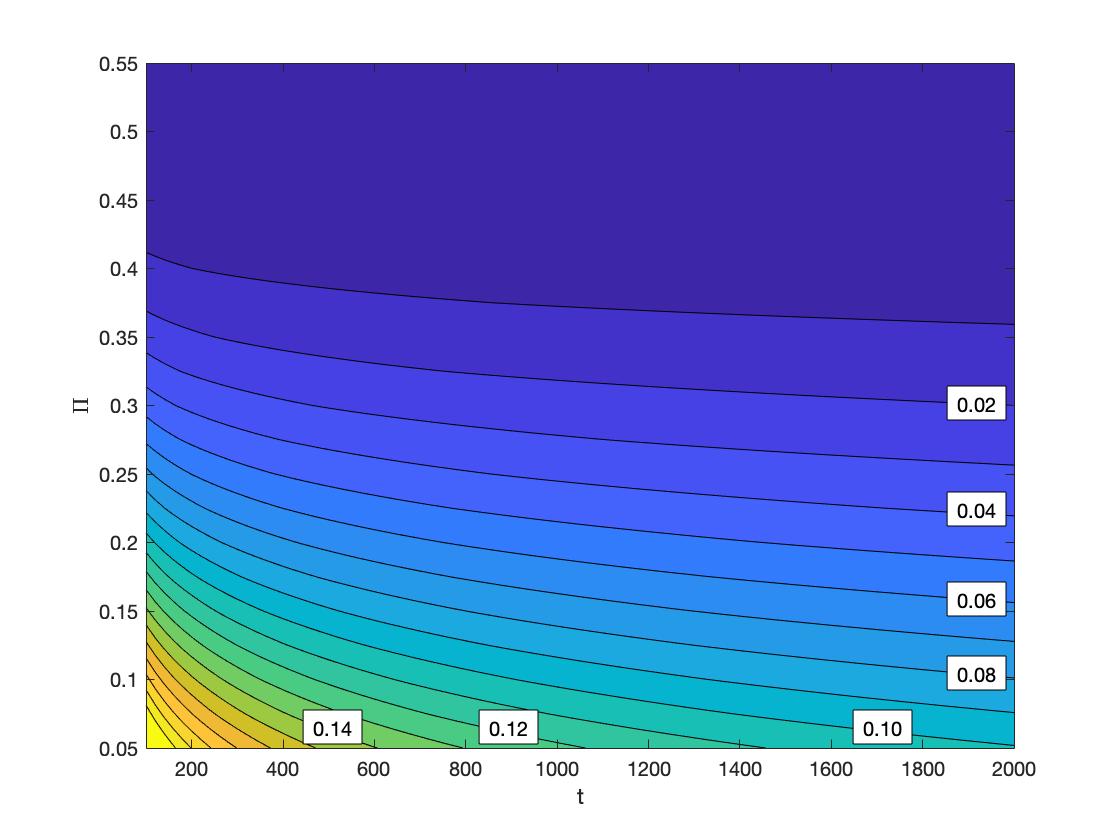}
\caption{ $w<6a$ }
\label{diagnostic}
\end{subfigure}
\begin{subfigure}{0.48\textwidth}
\includegraphics[width=\textwidth]{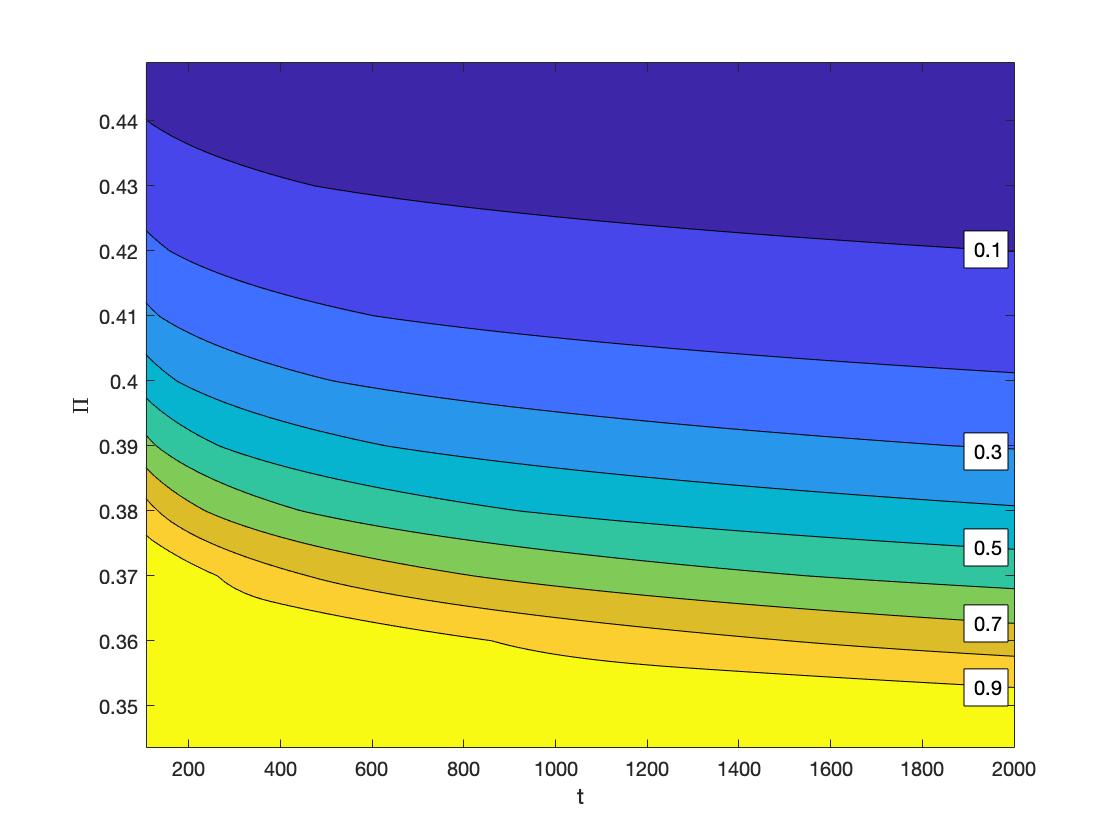}
\caption{Plug width $<2a$ }
\label{diagnostic2}
\end{subfigure}
\caption{Proportion of proppants by volume, predicted in fracture regions where $w<6a$, or plug width $<2a$, given average concentration, $\Pi$, and time, $t$.}
\end{figure}

Drawing on calculations we have made so far, we are now in a position to assess the significance of tip screen-out in our model, something we have neglected so far by adopting a continuum model of proppant transport. This is where, near the crack tip, the narrowing crack aperture causes proppants to jam and block the fracture, significantly affecting the development of the evolving formation and the convective transport of proppants. In \cite{donstovpierce} this problem is addressed using a `blocking function' which reduces proppant flux to zero in apertures smaller than three times the average particle's diameter. We will use this threshold to weigh the significance of ignoring screen-out in our model. Figure \ref{diagnostic} shows the volume-proportion of proppants predicted in fracture regions of width less than this threshold, dependant on the time, $t$, and the average proppant concentration, $\Pi$. We see that for early times and low concentrations, our model predicts a significant proportion of proppants in these regions, where the fracturing fluid is clear in reality. However, in concentrations greater than $0.3$ this proportion is relatively small; this means our model, which ignores tip screen-out, is self-consistent. This difference arises from the effective viscosity, which increases with $\Pi$ and causes the ratio of fracture width to length to decrease. 

Lecampion \& Garagash \cite{lecgara} conclude that their rheology, which is employed throughout this paper, agrees very well with experimental results when the predicted width of plug flow is greater than a particle's width. In figure \ref{diagnostic2}, we see this condition holds for moderate times when $\phi>0.4$. It does not for $\phi<0.4$. Therefore, in this regime we can expect slight mismatches between predicted and practical concentration profiles; this arises from a breakdown of the continuum model in the jammed part of the flow \cite{lecgara}. 

\section{Crack Closure: Problem Formulation}
\begin{figure}
\includegraphics[width=\textwidth]{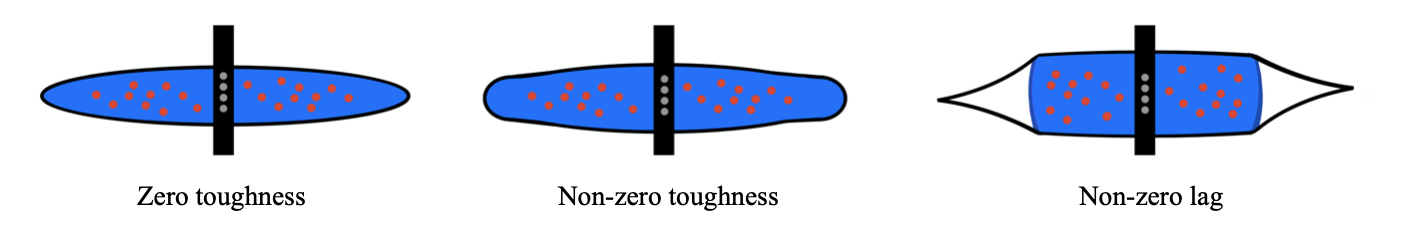}
\caption{Possibilities for modelling the crack tip.}
\label{schematics}
\end{figure}

In the zero-proppant case, Lai et al \cite{huppert} have confirmed experimentally that for late times after the fluid pressure is released, the crack radius is constant and volume scales as $t^{-1/3}$. It is tempting to repeat our previous work in order to find an asymptotic solution with a generalised total fracture volume $Qt^\alpha$. We would then let $\alpha=-1/3$ to model the case of closure. This approach leads us to
\begin{align}
\alpha \int_\rho^1 \xi h(\xi) d\xi +\beta \rho^2 h=- \frac{\rho h^3}{\eta_e}\pdv{p}{\rho},  \label{badegg}
\end{align}
in the place of (\ref{mueffeq}). Here $\beta=(3\alpha+1)/9 $ is the exponent for $L$, giving the radial growth of the fracture. However, we see that attempts to solve (\ref{badegg}) using the previous numerical method fail as $(\alpha,\beta)\to(-1/3,0)$, corresponding to the case in \cite{huppert}. This is because the tip asymptotes $w\propto (1-\xi)^{2/3}$ and $p \propto (1-\xi)^{-1/3}$ are a result of an \textit{advancing} fracture in a zero-toughness medium. Spence \& Sharp \cite{1985} note that $h\sim C(1-\xi)^\tau$ implies $p\sim C\tau (\cot \pi \tau) (1-\xi)^{\tau-1}$. Balancing terms in (\ref{badegg}), we are forced with $C\leq0$ if $\beta \leq 0$ which clearly can't lead to physical solutions, given the constraint $h\geq0$. In the same paper, solutions for $\beta=0$ are shown to exist without the assumption of zero-toughness; these have $h\sim(1-\xi^2)^{1/2}$. However, this causes difficulties in the case of an evolving fracture, since a non-zero toughness parameter, $\aleph$,  brings time dependence to the scaled equations we have derived. An alternative solution would be the addition of a non-zero fluid lag,  providing a region of negative pressure between the fluid front and the crack tip. Such a region exists in reality, containing either vapour from the fracturing fluid or, if the surrounding medium is permeable, pore fluid \cite{rubin,garalag}. Zero-toughness solutions using this formulation are explored in \cite{lag}. Schematics of each possible solution type are shown in Figure \ref{schematics}.

Any model utilising a time independent concentration profile is likely to fail in describing fracture closure at late times. This is because the width of the crack is decreasing as $t^{-1/3}$, so it is bound to become comparable to the proppant diameter. At the point where $\epsilon L/a\approx6$, the proppants begin to bridge across the fracture, effectively fixing them in position \cite{donstovpierce}; therein, concentrations will increase as the carrier fluid is forced from the cavity. For this reason, we will instead address the problem of finding the residual crack shape, given some axisymmetric initial distribution of proppants; we will assume these are radially immobile from the moment pressure is released. This method has been used with success to model the closure of a bi-wing fracture by Wang et al. \cite{wang,wang2}.

\section{Crack Closure: Residual Width Profiles}

We model the residual shape of the fracture using $w_{p}(r)$, defined as the close packed width of proppants. That is to say, after packing the proppants as tightly as possible in the z direction, so $\phi=\phi_{rcp}$, this is the residual width. Given some radial distribution of proppants described by the average concentration, $\Pi$, and un-scaled width profile, $w$, we deduce that $w_{p}=w\Pi/\phi_{rcp}$.  This description is compatible with the frictional rheology of Lecampion \& Garagash \cite{lecgara}, used previously, which asserts that a non-zero normal force on the proppants, along with vanishing shear stress, causes compression up to the random close packing limit. We then assume that the surrounding fracture simply collapses around the proppant pack. Our primary interest will be in using proppant distributions, arising from the injection phase described previously, to predict the geometry of the residual formation.

In \cite{wang2} a more complicated model is offered; this considers stress from the contact of opposing crack asperities, proppant embedment into the fracture walls, and compression of proppants. Since we will be concerned with cases where $w_p$ is non-zero along the entire crack radius; the contact term arising from the crack asperities, which is significant in the un-propped case, will not be necessary. Furthermore, in the same paper \cite{wang2} the depth of proppant embedment is shown to be of the order $K_e=a(3/4E')^2 (16mE'^2/9c_p)^{2/3}$. Here, $m\approx 2\sqrt 3$ is a constant which depends on the packing of proppants. Using the value of $c_p=3.9\times10^{-8} \textrm{Pa}^{-1}$ \cite{wang2}, as well as the typical values of $a=50\mu\textrm{m}$ and $E'=40\textrm{GPa}$ mentioned earlier, we note that $K_e\approx 1\mu\textrm{m} $, around 100 times smaller than the given proppant diameter. Since we will generally model proppant packs which are several times the size of the proppant diameter in width, we will ignore this phenomenon. Finally, we note that, according to our previous estimates, more than $10\textrm{s}$ into the injection phase we should expect pressures of less than $1\textrm{MPa}$. In \cite{wang2} the compressive stress required to reduce the width of the closely packed proppant bed from $w_p$ to $w$ is given by $1/c_p\ln(w_p/w)$; using this, the same stress would only cause a $4\%$ reduction in width. Since typical stresses involved in the closure phase are much smaller than this, we will model the proppants as incompressible. 

\begin{figure}
\centering
\begin{subfigure}{0.47\textwidth}
\centering
\includegraphics[width=\textwidth]{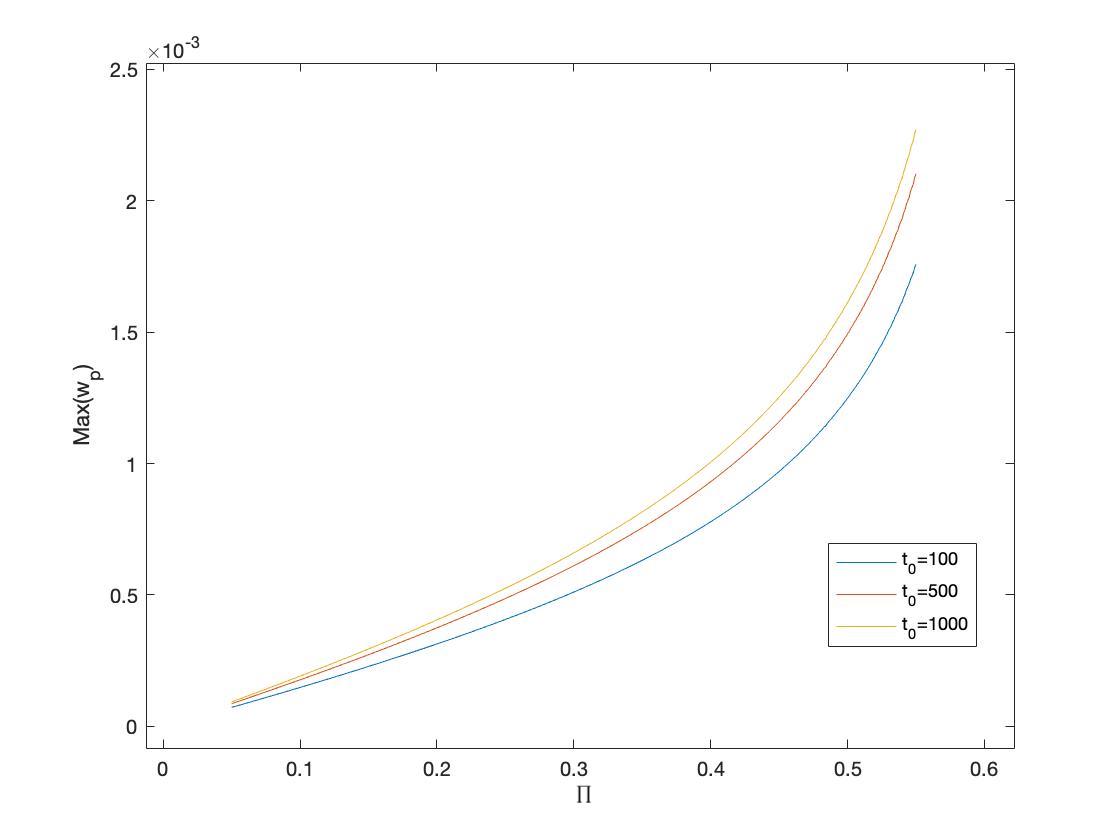}
\caption{Maximum fracture width.}
\label{maxwidths}
\end{subfigure}
\begin{subfigure}{0.47\textwidth}
\centering
\includegraphics[width=\textwidth]{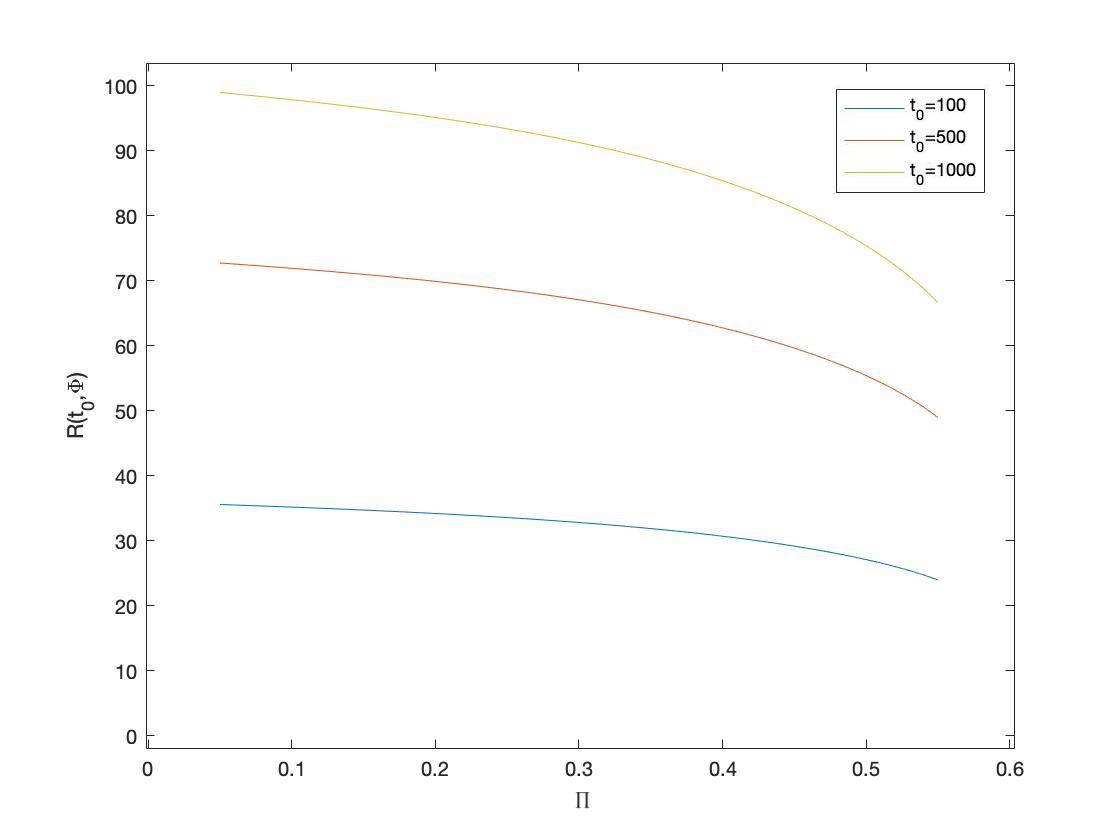}
\caption{Fracture radius.}
\label{finalrads}
\end{subfigure}
\caption{Plots showing the effect of average concentration on the maximum residual fracture width and radius for $t_0\in\{100,500,1000\}$.}
\end{figure}

This model of crack closure leads to a simple description of the residual crack profile. We have two parameters: one for average concentration, $\Pi$, and another for the time that injection ceases, $t_0$. Herein we will denote $\{\tilde h, \tilde p, \tilde \Gamma \}$ as the solution to the system of equations given in (\ref{absorbsystem}); $\tilde h$ and $\tilde p$ are plotted in Figure \ref{tildes} and we use the value $\tilde \Gamma  =0.598$. Then, using (\ref{effviscrels}) and the original scaling arguments, we deduce that
\begin{align}
w_p(\xi;t_0,\Pi)&=\frac{\Pi}{\phi_{rcp}}\epsilon(t_0)L(t_0)\eta_e (\Pi)^{2/9}\tilde \Gamma \tilde h(\xi), \\
R(t_0,\Pi)&=L(t_0) \eta_e (\Pi)^{-1/9}\tilde \Gamma.
\end{align}
From Figure \ref{tildes} we notice that $\max(\tilde h_1)\approx 1.35$. Using this, we may plot Figure \ref{maxwidths}, which shows the effect of average concentration on the maximum residual width of the formation. It is interesting to note that the propped width doesn't grow proportional to the proppant concentration, as one may expect from the close packing of the suspended proppants. Instead, the dependance is superlinear, because greater proppant concentrations lead to a higher effective viscosity; this causes the fracture to take a wider shape before the release of injection pressure. We can also see that $t_0$ has relatively little effect on the maximum crack width. This is because the $t_0$ dependent term, $\epsilon L$, grows with $t_0^{1/9}$. By contrast, in Figure \ref{finalrads} we see a greater time dependence in the final radius, which grows with $L\propto t^{4/9}$. As the proppant concentration increases, with $t_0$ fixed, we see a decrease in the final radius of fracture achieved, arising from an increase in the effective viscosity. 

\section{Conclusions}

We have established a mathematical framework that captures the behaviour of a slurry within a pressure driven cavity. Using typical parameters from industrial fracking, we predict that the development length, required to establish stable proppant flow away from the wellbore, is negligible compared to the typical radius of the penny-shaped fracture generated. As a result, we may assume the flow is fully developed, reducing the in-fracture distribution of proppants to a function of the radial distance from the wellbore. A further assumption of constant proppant injection rate allows us to describe the proppant distribution with one parameter, the total solid volume fraction. In the zero-concentration limit, our model becomes similar to one derived using Newtonian flow, with some disagreement arising from our choice of a dense frictional rheology. 

Within this framework, we are able to define an effective viscosity, which we may absorb into our equations using a suitable choice of scaling. This is a particularly striking result because it establishes an equivalence between slurry flow of a given solid fraction and simple Newtonian flow with some particular viscosity, at least in the sense of fracture development. Solving the resulting set of equations numerically, we may then return to our original scaling to investigate concentration-specific solutions. Unsurprisingly, we predict width and pressure profiles with the tip-asymptotic behaviour described in \cite{detsav}. As the proppant concentration increases we expect shorter and wider fractures with steeper fluid pressure gradients. In the centre of the fracture, where shear rate vanishes, we predict the formation of a disk of plug flow with width, in relation to the crack, increasing with the average proppant concentration. Evaluating our model, we see that the unaccounted effect of tip screen-out is likely to be significant in the low concentration, low effective viscosity case, particularly at early times. Here, the cavity formed is narrow, so near its tip, particle bridging is likely. Moreover, we observe that for typical fracturing timescales, if $\Pi<0.4$, our model predicts plug flow thinner than one particle width: suggesting that our use of a continuum model may not be appropriate. Otherwise, the plug flow is broader than a particle's width, meaning it is physically realisable and the results of \cite{lecgara} suggest we should have good experimental agreement.

Lastly, we have adopted a simple model of crack closure which regards the remaining proppants to be immobile and incompressible. This allows us to predict the shape of the residual crack, based on two parameters: the average proppant concentration within the injected fluid and the length of time between the initiation of fracking and the release of pressure. Simple formulae show that the residual fracture width increases significantly with proppant concentration, and grows very slowly with time; fracture radius however, decreases with proppant concentration and increases with time. 
							
The results established here have important applications in both contexts of industrial fracking and geological dike formation. Diagnostics of tip screen-out and forecasts of residual fracture geometry are relevant to the formation of conductive fractures, whilst predictions about the shape and particle distribution of a slurry driven crack relate more to a cooling magma. The discovery of an effective viscosity may also provide a foothold in understanding slurry driven fractures, particularly given the bounty of literature surrounding cracks generated by Newtonian fluid. In spite of all this, experimental investigation is necessary to bolster the predictions we have made. We hope this will form the basis of a second article, with tentative title: `Proppant flow in a penny-shaped crack. Part II : Experimental Investigation'. 

\section{Acknowledgements}
The authors would like to thank Derek Elsworth (Pennsylvania State University), Elisabeth Guazzelli (Centre National de la Recherche Scientifique) and Emmanuel Detournay (University of Minnesota) for their support and guidance in the drafting of this paper; with special gratitude to Elisabeth for providing the data used in Figure \ref{rheoplots}. We would also like to thank John Willis (University of Cambridge) for his support in the publication of the paper.


\appendix


\section{Integrating the \texorpdfstring{$\phi$,phi} conservation equation over the crack width}
\label{appendix1}
In this Appendix we integrate equation (\ref{prebigwidth}) over $(-w,w)$ to yield (\ref{bigwidth}); we will take a term-by-term approach. First, we note that by (\ref{fintroduction}),
\begin{align}
\int_{-z}^z \phi(\xi, \eta)d\eta&=2\int_{0}^z \mu^{-1}\left(\mu_w(\xi) \frac{\eta}{w}\right)d\eta, \\
&=2z\Pi\left(\mu_w(\xi) \frac{z}{w}\right). \label{Piintroduction}
\end{align}
Hence, we see that
\begin{align}
\int_{-w}^w \pdv{\phi}{\xi}d\eta&=\pdv{}{\xi}\int_{-w}^w \phi d\eta -2\phi(\xi,w)\pdv{w}{\xi}, \\
&=2\pdv{}{\xi}\left[w\Pi \circ \mu_w(\xi)\right] -2\phi(\xi,w)\pdv{w}{\xi}.
\end{align}
Then, integrating by parts, we find
\begin{align}
\int _{-w}^{w} \eta \pdv{\phi}{\eta} d\eta = 2\left[w\phi(\xi,w)-w\Pi \circ \mu_w(\xi)\right].
\end{align}
Furthermore, utilising the expression of $v_r$ given in (\ref{vrexpression}) and the condition $v_r(\xi,\pm w)=0$ we determine
\begin{align}
\int _{-w}^{w} \pdv{(\xi \phi v_r)}{\xi}d\eta&=\pdv{}{\xi} \left[ \xi \int_{-w}^{w}\phi v_r d\eta \right], \\
&=-\frac{6}{\Gamma}\pdv{}{\xi}\left[\xi \pdv{p}{\xi} \int_0^w \phi(\xi,\eta) \int_\eta^w \frac{I(\phi(\xi,z))z}{\mu(\phi(\xi,z)) } dzd\eta \right] ,\\
&=-\frac{6}{\Gamma}\pdv{}{\xi}\left[\xi \pdv{p}{\xi} \int_0^w \int_0^z \phi(\xi,\eta) \frac{I(\phi(\xi,z))z}{\mu(\phi(\xi,z)) } d\eta dz \right], \\
&=-\frac{6}{\Gamma}\pdv{}{\xi}\left[\xi \pdv{p}{\xi} \int_0^w z^2\Pi\left(\frac{\mu_w z}{w}\right) \frac{I(\phi(\xi,z))}{\mu(\phi(\xi,z)) } dz \right].
\end{align}
However, by (\ref{fintroduction}), $\mu(\phi(\xi,z))=\mu_w z/w$, so
\begin{align}
\int _{-w}^{w} \pdv{(\xi \phi v_r)}{\xi}d\eta&=-\frac{6}{\Gamma}\pdv{}{\xi}\left[\frac{w\xi}{\mu_w} \pdv{p}{\xi} \int_0^w z\Pi\left(\frac{\mu_w z}{w}\right) I\circ \mu^{-1}\left(\frac{\mu_w z}{w}\right) dz \right], \\
&=-\frac{6}{\Gamma}\pdv{}{\xi}\left[\frac{\xi w^3}{\mu_w(\xi)^2}\pdv{p}{\xi}\Omega \circ \mu_w(\xi)\right].
\end{align}
Finally, we know that
\begin{align}
\int_{-w}^w \pdv{(\phi v_z)}{\eta} d\eta=2\phi(\xi,w) v_z(\xi, w).
\end{align}
In the original scaling we have the boundary condition $v_z(x,w)=\pdv{w}{t} (x,t)$; in the lubrication scaling this becomes
\begin{align}
-\dot\epsilon L v_z(\xi, w) &= \left[\dot \epsilon L + \epsilon \dot L\right]w(\xi,T) -\epsilon L \xi \left[ \frac{\dot L}{L}+\frac{\Gamma' \dot T}{\Gamma} \right] \pdv{w}{\xi} +\dot T\pdv{w}{T}.
\end{align}
Hence,
\begin{align}
v_z(\xi,w)=\frac w3-\frac{4\xi}{3} \pdv{w}{\xi}, \label{vz}
\end{align}
and so
\begin{align}
\int_{-w}^w \pdv{(\phi v_z)}{\eta} d\eta=2\phi(\xi,w) \left[\frac w3 -\frac{4\xi}{3} \pdv{w}{\xi}\right].
\end{align}
Adding these terms together and making various cancellations, we derive equation (\ref{bigwidth}).


\section{Zero-Concentration Limit}

In this Appendix, we will compare the properties of equation (\ref{bigwidth2}) to the equivalent zero-proppant equation. Modelling the flow as Newtonian instead, we would have used the relation $\tau=\eta_f \dot\gamma$. In our choice of scaling this becomes $\tau=\dot \gamma$. Hence (\ref{normalstress}.2) is replaced by
\begin{align}
\pdv{v_r}{\eta}=\frac{3\eta}{\Gamma}\pdv{p}{\xi}, \label{newtonvr}
\end{align}													
where $\mathbf{v}$ is the fluid velocity. With the assumption that $\nabla \cdot v=0$, our scaled continuity equation is simply
\begin{align}
\frac{1}{\Gamma \xi} \pdv{(\xi v_r)}{\xi} +\pdv{v_z}{\eta}=0.
\end{align}
Integrating first over $(-w,w)$ as in Appendix \ref{appendix1}, making use of (\ref{vz}), (\ref{newtonvr}) and $\tau=\dot \gamma$, we obtain
\begin{align}
\frac w3-\frac{4\xi}{3} \pdv{w}{\xi}=\frac{1}{\xi\Gamma^2}\pdv{}{\xi}\left[\pdv{p}{\xi}\xi w^3\right].
\end{align}
Then, multiplying by $\xi$ and integrating from $\rho$ to 1, we use the $w^3\partial p /\partial \xi \to 0$ limit employed to derive (\ref{bigwidth2}),
\begin{align}
\int_\rho^1 \xi w d\xi + \frac49 \rho^2 w= -\frac{\rho w^3}{3\Gamma^2} \pdv{p}{\rho}. \label{fluideqn}
\end{align}

In order to compare (\ref{bigwidth2}) and (\ref{fluideqn}), we are required to find the limit of $\Omega/(x^2\Pi)$ as $x\to \infty$. Explicitly we see that
\begin{align}
\lim_{x\to \infty} \frac{\Omega(x)}{x^2\Pi(x)}&=\lim_{x\to \infty} \frac{1}{x^3\Pi(x)} \int_0^x \Pi(u) I \circ \mu^{-1}(u)udu, \\
&=\lim_{x\to \infty} \int_0^1 \frac{\Pi(vx)}{\Pi(x)} \cdot \frac{I\circ \mu^{-1}(vx)}{vx} \cdot v^2 dv, \\
&= \int_0^1v^2 \lim_{x\to \infty} \left[ \frac{\Pi(vx)}{\Pi(x)}  \right] dv, \label{jump1}\\
&=\int_0^1 v\lim_{x\to \infty} \left[ \frac{\int_0^{vx}\mu^{-1} (u)du}{\int_0^{x}\mu^{-1}(u) du} \right] dv, \\
&=\int_0^1 v^2\lim_{x\to \infty} \left[ \frac{\mu^{-1}(vx)}{\mu^{-1}(x)} \right] dv,  \label{lhopital}\\
&=\int_0^1 v^2\lim_{x\to \infty} \left[ \frac{I^{-1}(vx)}{I^{-1}(x)} \right] dv, \label{jump2}\\
&=\int_0^1 v^2\lim_{x\to \infty} \left[ \frac{1+\sqrt{x}}{1+\sqrt{vx}} \right] dv,\\
&=\int_0^1 v^{3/2} dv,\\
&=2/5.
\end{align}
Here (\ref{jump1}) and (\ref{jump2}) arise from the fact $I(\phi) \sim \mu(\phi)$ as $\phi \to 0$, because the fluid shear stress approaches the slurry shear stress. (\ref{lhopital}) comes from L'Hôpital's rule. We conclude that the equations governing Newtonian flow are not the same as those in the zero-proppant slurry flow limit. 
								

\section{Matrix \texorpdfstring{$(P)_{ij}$, when $N=4$}{Pij when N=4}}
The matrix $(P)_{ij}$ for $N=4$, as provided in \cite{detsav}, is given in table \ref{mij}.

\begin{table}[h]
{\renewcommand{\arraystretch}{1.15}
\begin{tabular}{|rr|rrrrrr|}
\hline
 &  &&  &  & j &&\\
 &  && -1 & 0 & 1 & 2 & 3 \\
 \hline
 & -1 && 1.0000 & 0.0000 & 0.0000 & 0.0000 & 0.0000 \\
 & 0 && 0.0000 & 0.9560 & 1.2730 & 0.4101 & 0.3145 \\
i & 1 && 0.0000 & 0.0991 & -0.0185 & 0.4068 & 0.0610 \\
 & 2 && 0.0000 & 0.0018 & -0.0429 & -0.0244 & 0.2293 \\
 & 3 && 0.0000 & 0.0017 & 0.0039 & -0.0416 & -0.0141 \\
 & 4 && 0.0000 & 0.0005 & 0.0026 & -0.0032 & -0.0372 \\
 \hline
\end{tabular}}
\caption{Matrix $(P)_{ij}$, for N=4.}
\label{mij}
\end{table}



\begin{thebibliography}{20}

\bibitem{history1}
Wells, Bruce A., ed. (2007). "Shooters". The Petroleum Age. American Oil and Gas Historical Society. 4 (3): 8–9. ISSN 1930-5915

\bibitem{history2}
Charlez, Philippe A. (1997). Rock Mechanics: Petroleum Applications. Paris: Editions Technip. p. 239. ISBN 9782710805861.

\bibitem{stressmeasure}
National Earthquake Hazards Reduction Program (U.S.), Geological Survey (U.S.), Office of Earthquakes, Volcanoes, and Engineering, U.S. National Committee for Rock Mechanics (1983). Hydraulic Fracturing Stress Measurements. Volume 26 of International journal of rock mechanics and mining sciences and geomechanics abstracts. 

\bibitem{geothermal}
Pierce, Brenda (2010). Geothermal Energy Resources. National Association of Regulatory Utility Commissioners (NARUC). 

\bibitem{CO2}
Miller, Bruce G. (2005). Coal Energy Systems. Sustainable World Series. Academic Press. p. 380. ISBN 9780124974517.

\bibitem{dikesummary}
E. Rivalta, B. Taisne, A.P. Bunger, R.F. Katz (2015). A review of mechanical models of dike propagation: Schools of thought, results and future directions. Tectonophysics. Volume 638,2015. Pages 1-42. ISSN 0040-1951.

\bibitem{magmatic}
Petford, N., Koenders, M.A. (1998). Granular flow and viscous fluctuations in low Bagnold number granitic magmas. Journal of the Geological Society, 155 (5), pp. 873-881. 10.1144/gsjgs.155.5.0873

\bibitem{1985}
Spence, D.A., Sharp, P.W. (1985). Self-similar solution for elastohydrodynamic cavity flow. Proc. Roy. Soc. London, Ser. A (400),289–313.

\bibitem{detsav}
A.A. Savitski, E. Detournay (2002). Propagation of a penny-shaped fluid-driven fracture in an impermeable rock: asymptotic solutions, International Journal of Solids and Structures, Volume 39, Issue 26, Pages 6311-6337.

\bibitem{einstein}
Einstein, A. (1906). A new determination of molecular dimensions. Ann. Phys. 4 (19), 289–306.

\bibitem{boyer}
Boyer F., Guazzelli É., Pouliquen O. (2011). Unifying suspension and granular rheology.
Phys. Rev. Lett. 107 (18), 188301.

\bibitem{comparative}
Dontsov EV, Boronin SA, Osiptsov AA, Derbyshev DY. (2019). Lubrication model of suspension flow in a hydraulic fracture with frictional rheology for shear-induced migration and jamming. Proc. R. Soc. A 475: 20190039.

\bibitem{lecgara}
Lecampion, Garagash (2014). Confined flow of suspensions modelled by a frictional rheology. J. Fluid Mech. (2014), vol. 759, pp. 197–235. Cambridge University Press 2014. doi:10.1017/jfm.2014.557


\bibitem{herbertpractical}
Niall J. O’Keeffe, Herbert E. Huppert \& P. F. Linden (2018). Experimental exploration of fluid-driven cracks in brittle hydrogels. J. Fluid Mech., vol. 844, pp. 435–458.



\bibitem{garadet}
Garagash, D.I., Detournay, E. (2000). The tip region of a fluid-driven fracture in an elastic medium. ASME J. Appl. Mech. 67, 183–192.


\bibitem{sneddon}
Sneddon, I.N., (1951). Fourier Transforms. McGraw-Hill, New York, NY


\bibitem{rice}
Rice, J.R., (1968). Mathematical analysis in the mechanics of fracture. In: Liebowitz, H. (Ed.), Fracture, an Advanced Treatise. Vol. II.Academic Press, New York, NY, pp. 191–311 (Chapter 3).


\bibitem{dagois}
Dagois-Bohy S., Hormozi S., Guazzelli É, Pouliquen O. (2015). Rheology of dense suspensions of non-colloidal spheres in yield-stress fluids. Journal of Fluid Mechanics, 776, R2. doi:10.1017/jfm.2015.329

\bibitem{garside}
Garside, J., Al-Dibouni, M. R. (1977). Velocity-voidage relationships for fluidization and
sedimentation in solid–liquid systems. Ind. Eng. Chem. Process Des. Dev. 16 (2), 206–214.


\bibitem{zaki}
Richardson, J., Zaki, W. (1954) Sedimentation and fluidization: Part I. Trans. Inst. Chem. Engrs
32, 35–47.


\bibitem{bacri}
Bacri, J.-C., Frenois, C., Hoyos, M., Perzynski, R., Rakotomalala, N. \& Salin, D. (1986). Acoustic study of suspension sedimentation. Europhys. Lett. 2 (2), 123–128.

\bibitem{shio}
Shiozawa, S., Mcclure, M. (2016). Simulation of proppant transport with gravitational settling and fracture closure in a three-dimensional hydraulic fracturing simulator. J. Petrol. Sci. Engng, 138, 298–314.


\bibitem{chen}
Chen Zhixi, Chen Mian, Jin Yan, Huang Rongzun (1997). Determination of rock fracture toughness and its relationship with acoustic velocity, International Journal of Rock Mechanics and Mining Sciences, Volume 34, Issues 3–4, 1997, Pages 49.e1-49.e11, ISSN 1365-1609

\bibitem{proppants}
Feng Liang, Mohammed Sayed, Ghaithan A. Al-Muntasheri, Frank F. Chang, Leiming Li (2016). A comprehensive review on proppant technologies. Petroleum, Volume 2, Issue 1, March 2016, Pages 26-39. 

\bibitem{donstov}
Dontsov EV, Boronin SA, Osiptsov AA, Derbyshev DY. (2019). Lubrication model of suspension flow in a hydraulic fracture with frictional rheology for shear-induced migration and jamming. Proc. R. Soc. A 475: 20190039.

\bibitem{abram}
Abramowitz, M., Stegun, I.A. (Eds.), (1964). Handbook of Mathematical Functions with Formulas, Graphs, and Mathematical Tables.Applied Mathematics Series, 55. US Govt. Print. Off, Washington, DC.

\bibitem{nelder}
Lagarias, J. C., J. A. Reeds, M. H. Wright, \& P. E. Wright (1998). Convergence Properties of the Nelder-Mead Simplex Method in Low Dimensions. SIAM Journal of Optimization. Vol. 9, Number 1, 1998, pp. 112–147.

\bibitem{donstovpierce}
Dontsov, E. V., Peirce, A. P. (2014). Slurry flow, gravitational settling and a proppant transport
model for hydraulic fractures. J. Fluid Mech. 760, 567–590.



\bibitem{huppert}
Ching-Yao Lai, Zhong Zheng, Emilie Dressaire, Guy Z. Ramon, Herbert E. Huppert, \& Howard A. Stone (2016). Elastic Relaxation of Fluid-Driven Cracks and the Resulting Backflow. Physical Review Letters 117, 268001.

\bibitem{rubin}
A.M. Rubin. (1993). Tensile fracture of rock at high confining pressure: implications for dike propagation. J. Geophys. Res., 98 (B9) (1993), pp. 15,919-15,935. 

\bibitem{garalag}
E. Detournay \& D. Garagash (2003). The tip region of a fluid-driven fracture in a permeable elastic solid. J. Fluid Mech., 494, pp. 1-32.

\bibitem{lag}
D. Garagash (2006). Propagation of a plane-strain hydraulic fracture with a fluid lag: Early-time solution, International Journal of Solids and Structures 43, 5811–5835.

\bibitem{wang}
Jiehao Wang, Derek Elsworth \& Martin K. Denison (2018). Propagation, proppant transport and the evolution of transport properties of hydraulic fractures. J. Fluid Mech., vol. 855, pp. 503–534.


\bibitem{wang2}
Wang, J. \& Elsworth, D. (2018). Role of proppant distribution on the evolution of hydraulic fracture
conductivity. J. Petrol. Sci. Engng 166, 249–262.

\end{thebibliography}
\end{document}